\documentclass[a4paper,10pt]{article}

\usepackage[utf8]{inputenc}
\usepackage[T1]{fontenc}
\usepackage{graphicx}
\usepackage{xr-hyper}
\usepackage{hyperref}       
\usepackage{url}            
\usepackage{a4wide}
\usepackage{mathtools}
\usepackage[table,dvipsnames]{xcolor}         
\usepackage{mathtools}
\usepackage{authblk}

\definecolor{Magenta}{rgb}{0.8, 0.1, 0.6}

\hypersetup{colorlinks = true, linkcolor = NavyBlue,
            urlcolor  = gray,
            citecolor = NavyBlue,
            anchorcolor = NavyBlue}

\title{Job loss disrupts individuals' mobility and their exploratory patterns}

\author[1,+,*]{Simone Centellegher}
\author[1]{Marco De Nadai}
\author[1,2]{Marco Tonin}
\author[1]{Bruno Lepri}
\author[1,3,4,+,*]{Lorenzo Lucchini}
\affil[1]{Fondazione Bruno Kessler (FBK), Trento, Italy}
\affil[2]{Department of Sociology and Social Research, University of Trento, Trento, Italy}
\affil[3]{Centre for Social Dynamics and Public Policy, Bocconi University, Milan 20100, Italy}
\affil[4]{Institute for Data Science and Analytics, Bocconi University, Milan 20100, Italy}

\affil[*]{centellegher@fbk.eu, llucchini@fbk.eu}

\affil[+]{these authors contributed equally to this work}

\begin{document}

\flushbottom
\maketitle
\vspace{-0.5cm}
\vspace{0.5cm}

\begin{abstract}
In recent years, human mobility research has discovered universal patterns capable of describing how people move. These regularities have been shown to partly depend on individual and environmental characteristics (e.g., gender, rural/urban, country).
In this work, we show that life-course events, such as job loss, can disrupt individual mobility patterns. Adversely affecting individuals' well-being and potentially increasing the risk of social and economic inequalities, we show that job loss drives a significant change in the exploratory behaviour of individuals with changes that intensify over time since job loss.
Our findings shed light on the dynamics of employment-related behavior at scale, providing a deeper understanding of key components in human mobility regularities. These drivers can facilitate targeted social interventions to support the most vulnerable populations.
\end{abstract}


\section*{Introduction}
Economic and human behavioural statistics are crucial for effective decision-making. Large-scale population surveys have been invaluable in observing economic shocks and their implications. For example, unemployment data serve as a vital indicator of an economy's health and performance \cite{bls_unemployment}: when workers become unemployed, it affects their well-being and that of their families, it diminishes their purchasing power, and impacts the overall economy. 
However, conventional methods to track unemployment and its implications have been challenged by survey participation rates decline~\cite{krueger2017evolution, ons2023labour} especially in developing countries~\cite{international2015world,dewan2022world}.

Recently, a transformative shift has emerged through the utilization of large-scale behavioural data collected from technologies like mobile phones, GPS trackers, social media platforms, and credit cards. This shift have been instrumental in advancing research in human mobility \cite{gonzalez2008understanding,pappalardo2015returners,lucchini2019following,alessandretti2020scales}, financial well-being and purchase behaviour \cite{singh2015money,tovanich2021inferring,dong2017social,matz2016money,lucchini2022reddit,sobolevsky2016cities}, segregation and economic inequalities \cite{wang2018urban,chetty2022sociala, chetty2022socialb, moro2021mobility,yabe2023behavioral}, crime \cite{wang2016crime,song2019crime,luca2023crime} and public health \cite{wesolowski2012quantifying,oliver2020mobile,kraemer2020effect,aleta2020modelling,luca2023crime}.

Few studies in human mobility research have managed to estimate job loss at a fine-grained level ~\cite{moriwaki2020nowcasting,sundsoy2016estimating,toole2015tracking} with even fewer studies focused on the behavioural specificities of unemployed individuals~\cite{toole2015tracking,almaatouq2016mobile,barbosa2021uncovering}. However, the impact of job loss on individual mobility behavior at scale still remains a largely unexplored area of study.

In this context, the contribution of our work is twofold. Firstly, we introduce a real-time methodology for inferring unemployment status based on individual GPS trajectories. Secondly, we provide evidence of significant changes in mobility behaviour regularities following a job loss, particularly affecting vulnerable groups already at an increased risk of segregation \cite{moro2021mobility}.

We leverage a dataset of privacy-enhanced longitudinal GPS mobility traces of nearly 1 million anonymous opted-in individuals from January 3, 2020, to September 1, 2020, across several US states.
In order to preserve privacy, the data provider obfuscates devices' home locations to the Census Block Group level and removes visits to sensitive points of interest from the dataset.
The states are selected based on their diverse workforce composition profiles, enabling us to estimate unemployment at scale and analyze multiple facets of individuals' mobility behaviour following job loss. 
To ensure the representativeness of the GPS data and address potential sample biases \cite{yabe2023behavioral}, we employ a reweighting technique. This process generates a resampled cohort that reflects the demographic characteristics and the employed workforce across industrial sectors in all the states under study. We evaluate our methodology in the context of the COVID-19 pandemic, discussing its versatility for more general systemic shocks. 

Our analysis sheds light on the impacts of job loss, providing a comprehensive, multidimensional view of individuals' mobility patterns. This includes their geographic displacement, time allocation, and set of visited locations. We also show, through a temporal-independent analysis of employed versus unemployed behavioural patterns, that there is an increasing disparity in mobility behaviour between employed and unemployed individuals since the time of job loss. In this perspective, we also illustrate how demographic factors such as sex, age, income, race, and education level can intensify the impact of job loss on individual mobility pattern contraction.

Overall, our results provide evidence for the long-term effects of unemployment on individuals' daily lives. Job loss, as a major event in an individual's life, not only perpetrates but also exacerbates existing socio-demographic disparities in mobility behaviours. While prior literature on human mobility has identified universal characteristics in mobility patterns~\cite{noulas2012tale,alessandretti2018evidence,schlapfer2021universal,barbosa2018human}, our findings highlight that individual life-course events, such as job loss, can affect these regularities at the individual level. Such events have the potential to influence people's habits, as well as their social and psychological well-being.

\section*{Results}

\subsection*{Inferring individual employment status}
\begin{figure}[!ht]
\centering
\includegraphics[width=0.9\linewidth]{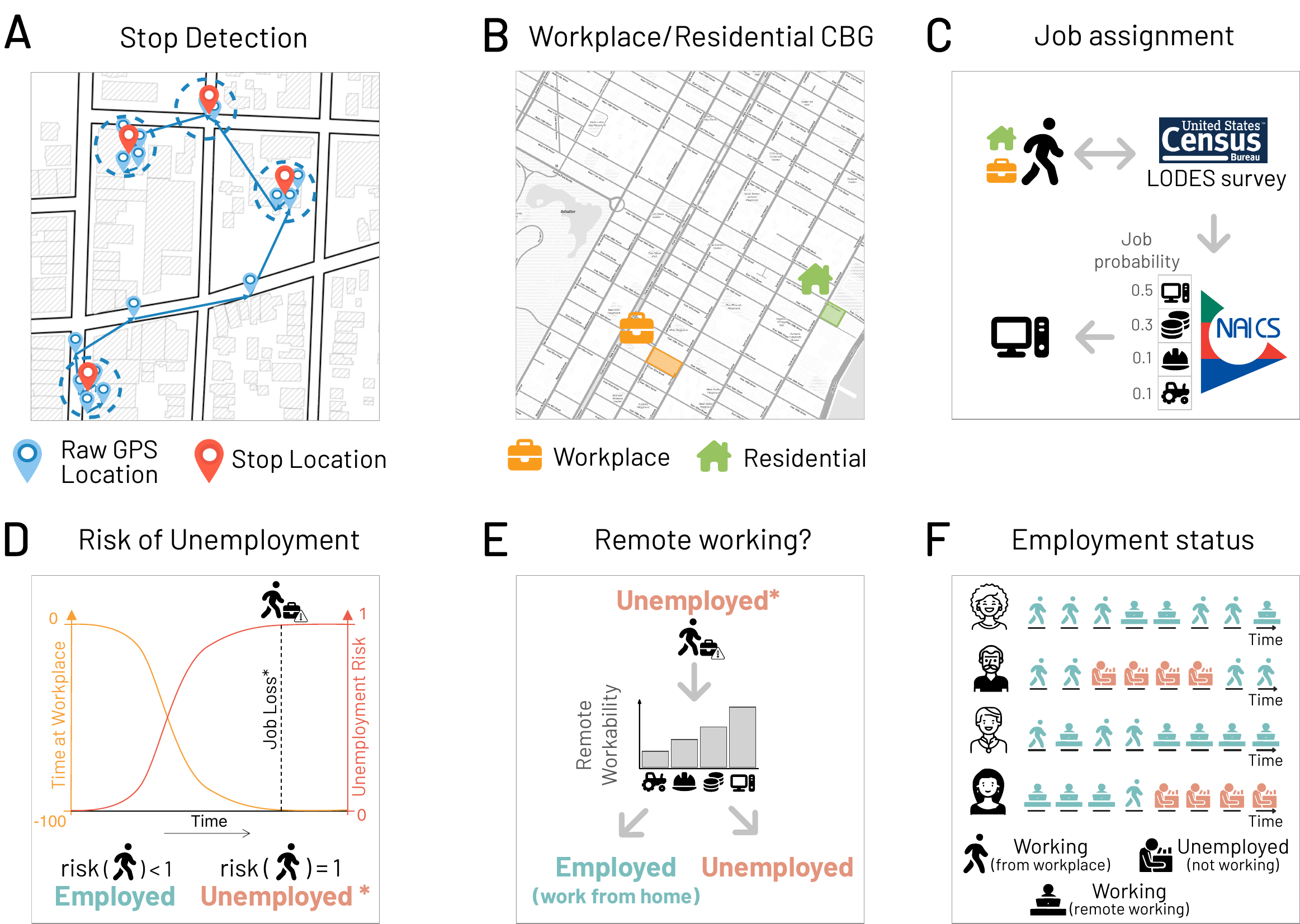}
\caption{\textbf{Employment status detection algorithm. } Overview of the developed procedure to detect unemployment. \textbf{(A)} Stop locations detection (individual stopped in a 65-meter radius and stayed for at least 5 minutes); \textbf{(B)} Workplace and Residential Census Block Groups (CBGs) detection; \textbf{(C)} Job Assignment, a NAICS sector is assigned in probability given the individual's Workplace and Residential CBGs; \textbf{(D)} the Risk of Unemployment is computed for each individual based on their workplace visits; \textbf{(E)} Remote working correction based on the teleworkability of the individual job (NAICS sector); \textbf{(F)} For each individual we have their full employment status over time. Icons: Fontawesome, Flaticon, Maps: Stamen Maps.}
\label{fig:step_by_step}
\end{figure}

To determine an individual's employment status, we have devised a procedure that integrates both location and survey data (see Fig. \ref{fig:step_by_step}).
We use a large and longitudinal dataset of privacy-enhanced GPS location data collected across seven US states from January to September 2020 to enrich census survey data provided by the US bureau. Particularly, we use the Longitudinal Employer-Household Dynamics (LEHD) Origin-Destination Employment Statistics (LODES)~\cite{LODES}, which provides privacy-preserved statistics about the US workforce divided by industrial sectors classified with the North American Industry Classification System (NAICS)~\cite{NAICS} and provides data about how many individuals are employed in a specific NAICS sector, based on their census block groups (CBGs)~\cite{CBG} of residence and workplace. We refer to the Methods section for further details.

For each individual, we first identify stop locations, defined as sequences of GPS coordinates within a 65-meter radius where a user stayed for a minimum of 5 minutes (Fig. \ref{fig:step_by_step}A). Then, after detecting the individual's residential and workplace locations (Fig. \ref{fig:step_by_step}B), we enrich these locations with the LODES data information.

For those individuals with a detected work location, we label them as employed during the period in which their workplace location is identified. Each of these individuals is further assigned in probability a job, more specifically, a NAICS sector based on available survey data looking at their residential and workplace CBGs (Fig. \ref{fig:step_by_step}C). 

As a following step, individuals are labelled as ``at risk of unemployment'' based on the reduction in visits to the workplace: if individuals never visit their workplace location, they are considered as potential candidates for unemployment (Fig. \ref{fig:step_by_step}D). The determination of employment status is sampled taking into account both the risk status and the NAICS-specific likelihood of working from home at any given time.
 
To account for whether an individual is working from home, we leverage information on the ``teleworkability'' of jobs, as presented in the study of Dingel and Neiman \cite{dingel2020many}. For each industrial sector, the data provides the percentage of work that can be performed remotely (Fig. \ref{fig:step_by_step}E).
Based on the individual's job sector (NAICS), population-wide change in the time spent at work, and the weight of each individual in contributing to that change, we infer the unemployment status over time, thus determining whether the individual is working from home or is unemployed (see Methods section). 

\subsection*{Cohort selection and algorithm evaluation}
The privacy-enhanced location data provided by the location intelligence company Cuebiq intentionally excludes any direct information about users' employment to safeguard privacy. This absence of direct job-related information presents challenges in establishing ground truth for individuals’ employment status. Consequently, we evaluate the accuracy of our methodology at an aggregate level by leveraging aggregated monthly statistics from Unemployment Insurance (UI) claims and Local Area Unemployment Statistics (LAUS) datasets (see SI S1 for dataset details). UI claims data offer near real-time information on the number of claimants, reported weekly, providing a timely basis for our algorithm evaluation. In contrast, LAUS data represent official unemployment figures derived through an estimation process that incorporates multiple sources, including UI claims, but are published and consolidated with a longer delay. Additionally, we incorporate state-level employment information from the Bureau of Labor Statistics (BLS) through the Quarterly Census of Employment and Wages (QCEW) program (see SI S1 for additional data details).

All our analyses are conducted on a cohort of mobile phone users residing in the US. We employ individual reweighting to reconstruct a representative cohort sample that accurately mirrors both (i) population-wide representativeness, based on census block group (CBG) population data, and (ii) representativeness of the employed workforce population in each state across various industrial NAICS sectors, drawing from state-level employment statistics (BLS statistics). This post-stratification procedure is crucial for addressing potential biases within the location data and ensuring the data's representativeness~\cite{yabe2023behavioral, lucchini2023socioeconomic}. It enables us to compare employment status with Unemployment Insurance claims and LAUS unemployment figures and subsequently examine mobility patterns at a population-wide scale. Further details on the post-stratification technique can be found in SI S2A.

Our study focuses on seven U.S. states: New York, Wyoming, Indiana, Idaho, Washington, North Dakota, and New Mexico. These states were selected to capture diverse workforce compositions spanning primary, secondary, and tertiary economic sectors, as well as geographic diversity across the U.S. (see SI S2B for more details).

To assess the reliability of our job detection methodology, we use both LAUS and UI claims data. The UI claims enable near real-time evaluation of the algorithm at the monthly level, segmented by NAICS sectors. At the state level across the seven studied states, we observe a Pearson correlation coefficient of 0.89 between the monthly rate of individuals detected as unemployed (reweighted to match the employed population) and the monthly UI claims rate. Using the official unemployment statistics from the LAUS data, we find a Pearson correlation of 0.72 at the state level and 0.53 at the county level. These results demonstrate that our algorithm is fairly reliable and can estimate unemployment at an aggregate level (see SI S4 for further details on the algorithm evaluation).

\subsection*{Behavioural disparities between employed and unemployed individuals}
The availability of inferred individual employment status data over time provides a unique opportunity to gain insights into the impact of job loss on human mobility. The wealth of information at our disposal allows us to characterize and quantify changes in behaviour by comparing the daily mobility of individuals identified as employed or unemployed, offering a better understanding of shifts in mobility patterns following a job loss. In this study, we address two key questions: (i) How did individuals who experienced a job loss navigate through the pandemic period?; And, more broadly, (ii) what are the effects of job loss on an individual's mobility behaviour, and what happens when individuals face a prolonged period of unemployment?

To address these questions and ensure a fair comparison between a population of employed individuals and a population of unemployed individuals, we exclude all stop locations associated with an individual's workplace from our analysis. Therefore, our analysis focuses on extra-work individuals' mobility patterns, with a specific focus on those individuals who had been employed (even briefly) between January 3rd, 2020, and March 7th, 2020, namely before the WHO declaration of the COVID-19 pandemic (March 11th, 2020).

Considering systemic external factors in our analysis and given the significant stress placed on the labour market by the pandemic, we have the ideal conditions to study the consequences and gain a comprehensive understanding of the effects of job loss on individuals' mobility behaviour.

\begin{figure}[!htb]
\centering
\includegraphics[width=0.9\linewidth]{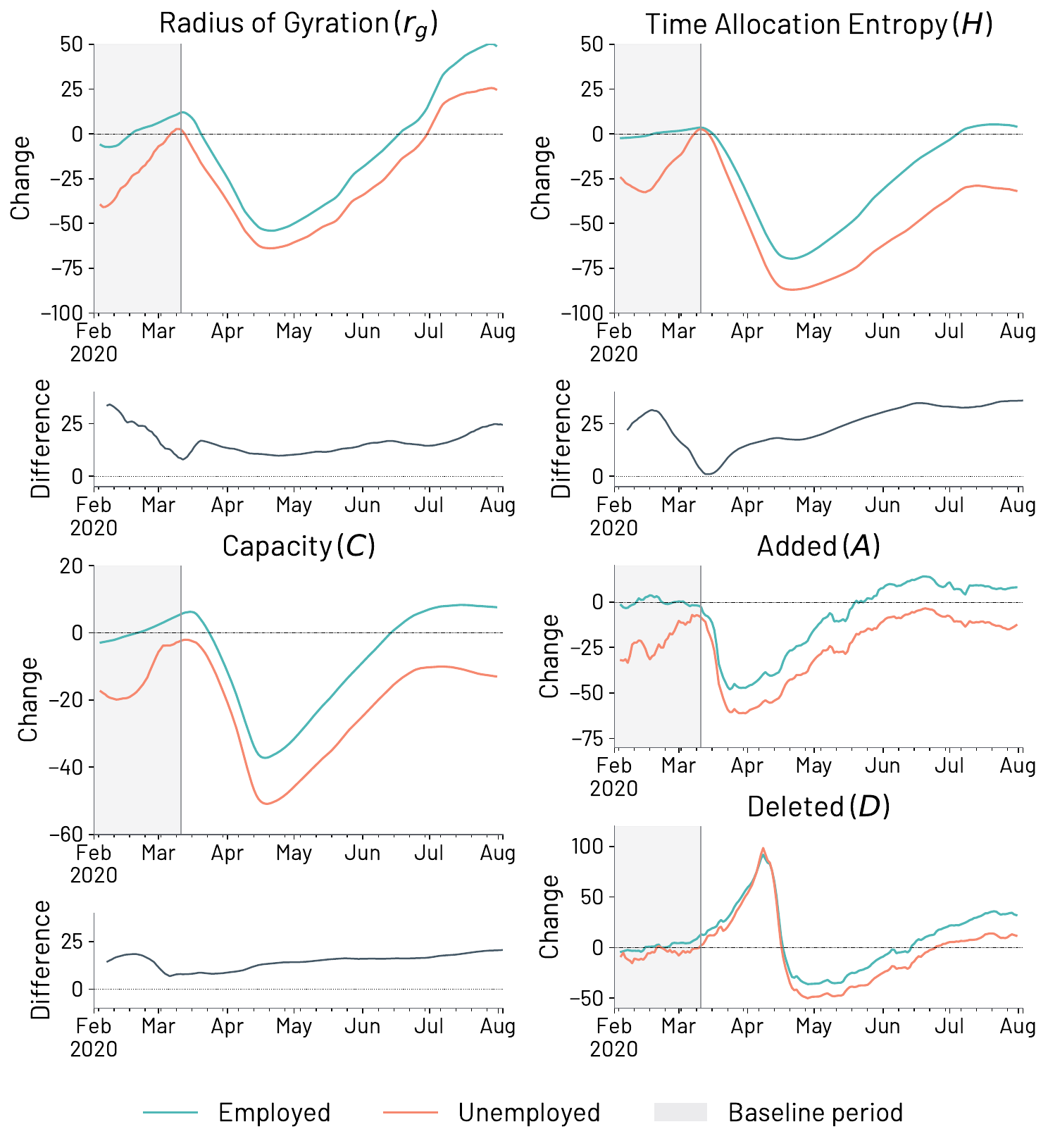}
\caption{\textbf{Impact of the pandemic on employed and unemployed mobility. } Percentage changes with respect to the baseline period (February 1st - March 7th) in extra-work individuals' mobility patterns for employed and unemployed groups, and their difference over time, as measured by different mobility metrics. \textbf{(A)} Radius of gyration ($r_g$) and the corresponding difference between the groups of employed and unemployed over time; \textbf{(B)} Time allocation entropy ($H$), which measures the distribution of time allocation in each visited location; \textbf{(C)} Capacity $C$ which represents the number of a user's familiar locations; and \textbf{(D)} the number of added $A$ and deleted $D$ locations between consecutive windows.
Each mobility metric is computed over a window of 28 days with a 1-day shift. The grey vertical line represents the WHO COVID-19 pandemic declaration (March 11th, 2020). 
Mobility patterns are reported only for individuals visiting their work location at least once within the grey shaded area period.}
\label{fig:emp_vs_unempl}
\end{figure}

Note that, as previously explained, our analysis focuses on individuals who were employed, even briefly, before the pandemic. Therefore, the curves representing the mobility indicators for unemployed individuals in the baseline period may not be representative. 

\subsubsection*{The disproportionate impact of the pandemic on unemployed individuals' mobility}

To provide a comprehensive analysis of changes in mobility, we measure within-individual variations by comparing activities to a baseline period preceding the pandemic (February 1st, 2020 - March 7th, 2020). Our focus revolves around three key well-known mobility metrics: (1) the radius of gyration ($r_g$) \cite{gonzalez2008understanding}, which measures the characteristic geographical displacement of individuals; (2) the time allocation entropy ($H$) \cite{song2010limits}, which measures the distribution of time allocation in each visited location; and (3) the users' locations' capacity, denoted as $C$, which captures the number of a user's familiar locations, alongside the number of locations added ($A$) to, and deleted ($D$) from the set of familiar locations within a specific time interval \cite{alessandretti2018evidence}  (find the formal definition of the mobility metrics in the ``Methods'' section). Collectively, these measures offer a multidimensional perspective on both the characteristic displacement and the complexity of individuals' exploratory behaviour.

In Fig. \ref{fig:emp_vs_unempl}, we present the results over time for each of these metrics and the relative difference over time between the group of employed and unemployed individuals. We consistently compare the mobility patterns of inferred unemployed individuals with those of employed individuals under similar pandemic-related conditions and restrictions. This approach helps isolate the specific effects of unemployment from broader external factors, such as lockdown measures. All the mobility metrics are computed for a window of 28 days with a 1-day shift. Note that, as previously explained, our analysis focuses on individuals who were employed, even briefly, before the pandemic. Therefore, the curves representing the mobility indicators for unemployed individuals in the baseline period (grey-shaded area) are not informative due to the low number of unemployed individuals during that period.
The results shown in Fig. \ref{fig:emp_vs_unempl} reveal a substantial impact of the pandemic on individual mobility patterns, particularly among unemployed individuals. Notably, the group of unemployed individuals exhibits lower overall activity levels across all the mobility metrics under study. Moreover, we observe that as the pandemic progresses, the mobility gaps between the employed and unemployed groups widen. While the reduction in mobility for the unemployed is limited when examining the individuals' characteristic displacement measured by the radius of gyration, with employed individuals reaching a low point of $-54\%$ and unemployed individuals reaching a low point of $-64\%$, the same is not true when looking at regularity and exploration patterns. The drop in activity is particularly pronounced for unemployed individuals when examining the time allocation entropy, with low points of $-70\%$ and $-87\%$ for employed and unemployed individuals respectively, and capacity, with low points of $-37\%$ and $-51\%$ for employed and unemployed individuals respectively.

From this analysis, it becomes evident that the routinary and exploratory behaviours of unemployed individuals, as measured by the time allocation entropy and by the capacity (together with its location turnover of added $A$ and deleted $D$ locations), were more affected than those of employed individuals. Moreover, over time, there is an evident increasing trend in the difference between the behaviours of the two groups. The difference between the two groups at the end of the period under study is $24\%$ for the radius of gyration, $36\%$ for the time allocation entropy and $21\%$ for the capacity.
Interestingly, following the gradual reduction of COVID-19 restrictions, there appears to be a clear (partial) recovery for all the different facets of mobility behaviour we analyzed.

\begin{figure}[!htb]
\centering
\includegraphics[width=\linewidth]{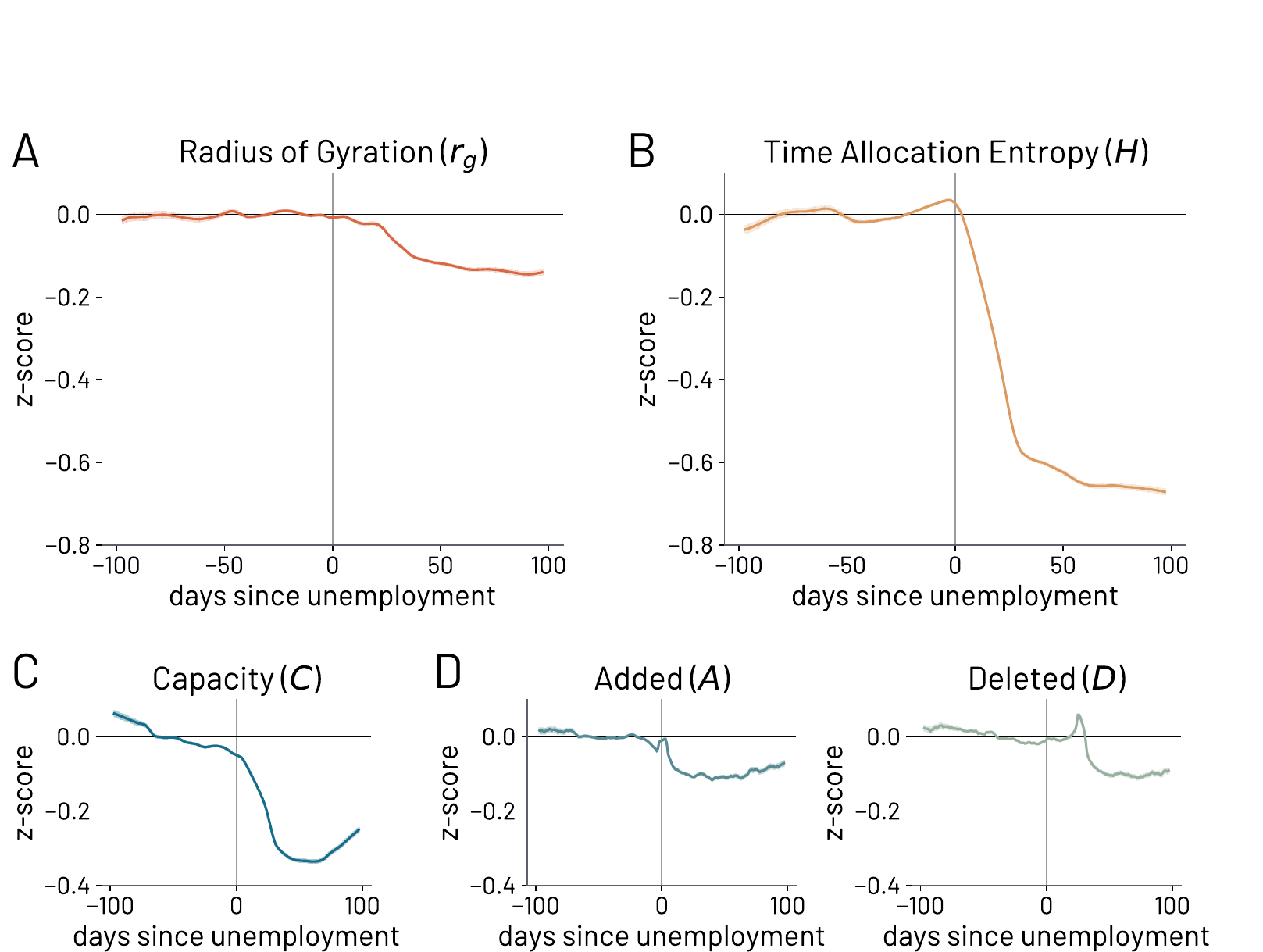}
\caption{\textbf{Individual-level mobility behaviour after job loss using employed population as reference group.} The results show the lasting impact of prolonged periods of unemployment on individual-level mobility behaviour.
Each individual's mobility indicator, which includes \textbf{(A)} the radius of gyration, \textbf{(B)} the time allocation entropy, \textbf{(C)} the capacity $C$, and \textbf{(D)} the added $A$ and deleted $D$ locations over time, is standardized by calculating the z-score using the average and standard deviation of the employed group's indicators on a specific day $t$. Then, time is aligned such that at time $t=0$, individuals have lost their jobs. Shaded areas represent the 2-standard-deviation range.}
\label{fig:after_job_loss}
\end{figure}

\subsubsection*{Prolonged unemployment and the deterioration of mobility behaviour} \label{sec:prolonged_unemp}
In the previous section, we provided insights into the collective mobility dynamics of employed and unemployed individuals, uncovering a disproportionate mobility response during the pandemic period between the two groups. To extend the validity of our findings beyond the pandemic conditions and ensure their generalizability to other possible systemic shocks, we further investigate into the growing divergence over time between the mobility behaviours of employed and unemployed individuals. Through the following analysis, we aim to understand the effects of job loss on individual-level mobility behaviour, assessing whether a prolonged period of unemployment leaves a lasting impact.

Hence, we present a robust and general framework for detecting and tracking unemployment potentially adaptable to different systemic shocks. In particular, we propose a time-independent analysis of employed/unemployed behavioural patterns which tries to understand whether the duration of unemployment contributes to the growing disparity between the two groups' mobility behaviour.

Due to the period during which the data was collected, we first need to consider the non-negligible impact of Non-Pharmaceutical Interventions, and more in general of the pandemic, on the general population mobility during 2020. To mitigate the effect of the COVID-19 pandemic on the results, we standardize each individual's mobility indicator by calculating the z-score using the average and standard deviation of the employed group's indicators on a specific day $t$. Then, to better understand the effects of a job loss on an individual, we align the mobility indicators of all individuals by shifting time so that $t=0$ represents the time when an individual lost their job. This approach enables consistent comparisons of individuals' mobility behaviour at different times with respect to the date at which they lost their jobs.

As illustrated in Fig. \ref{fig:after_job_loss}, both the radius of gyration (at a smaller level) and time allocation entropy (at a larger level) were affected and gradually decreased over time, reaching almost $-0.15$ and $-0.8$ standard deviations, respectively, compared to when individuals were employed. Although the radius of gyration seemed to be less affected, the relative time allocation entropy of individuals who lost their jobs decreased sharply and constantly the longer they were unemployed. This large reduction in time allocation entropy may be related to the tendency of unemployed individuals to spend a significant fraction of time at home \cite{almaatouq2016mobile, llorente2015social}.

A similar dynamic is observed in the capacity $C$ of individuals, which displays a sharp decrease of more than $-0.3$ standard deviations, followed by a slow recovery after approximately 60 days. The added locations ($A$) to the set of familiar places exhibited similar behaviour as the capacity, with a noticeable decrease after an individual loses their job. On the other hand, the deletion of familiar locations ($D$) increases abruptly when individuals lose their jobs, followed by a sharp decrease. Both the added and deleted locations then remain significantly low, indicating an overall lower turnover in the set of an individual's familiar locations. Despite a modest recovery after approximately two months, the results highlight the clear and persistent impact of unemployment on limiting individuals' abilities to explore new opportunities in physical space.
The drop in capacity ($C$) together with the decrease in the number of added ($A$) and deleted ($D$) locations highlight a reduced location turnover and a sustained contraction into the individuals' set of familiar locations. For an individual, this scenario may indicate a potential decrease in exposure to opportunities and an increased risk of isolation after experiencing a job loss.

\subsection*{Socio-demographic factors in job loss behavioural changes}
To get a better understanding of the implications of prolonged unemployment, we evaluate and quantify socio-demographic differences in mobility patterns among individuals enduring a prolonged period of unemployment. 
Leveraging socio-demographic information from the Longitudinal Employer-Household Dynamics (LODES) dataset \cite{LODES}, including \textit{Sex}, \textit{Age}, \textit{Income}, \textit{Race}, and \textit{Education}, we analyze demographic differences in mobility behaviours. 

Building on the results presented in Fig.~\ref{fig:after_job_loss}, we disaggregate the mobility behaviour of unemployed individuals ($t > 0$) based on the individual's socio-demographic group (see SI S5 for more details), revealing significant disparities in the mobility behaviour of unemployed individuals when compared with the mobility behaviour of employed individuals (see SI Tab. S3).

\begin{figure}[!ht]
\centering
\includegraphics[width=0.98\linewidth]{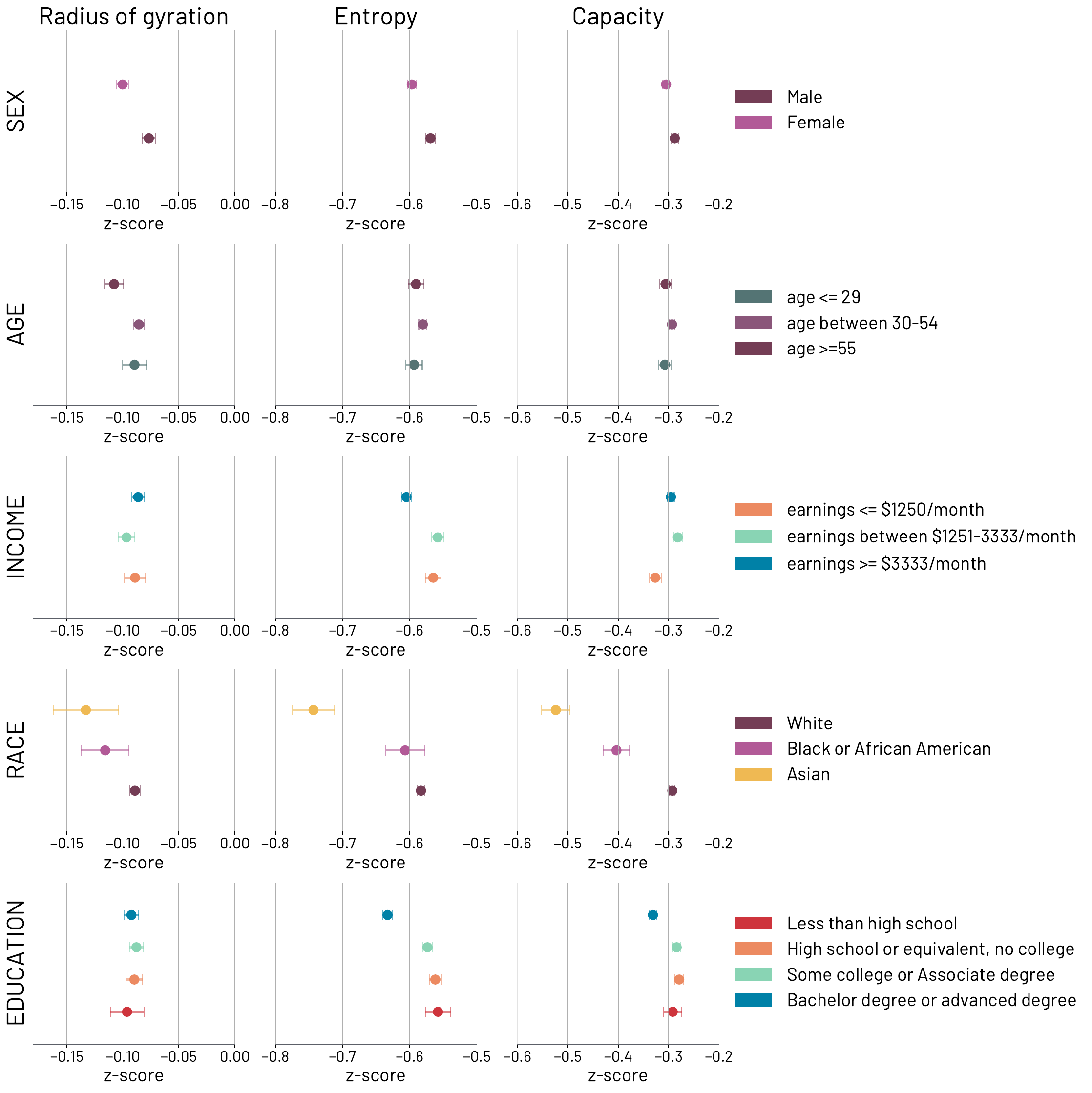}
\caption{\textbf{Demographic variations in mobility among individuals enduring a prolonged period of unemployment. } Differences in mobility behaviour of unemployed individuals in the latter stages of unemployment (between 30 and 100 days) for Sex, Age, Income, Race and Education demographics compared to the mobility indicators of the reference group of employed individuals. For each metric and demographic group, we provide the mean and standard error of each group in the 30-100 days period after the job loss.}
\label{fig:demographic_diff}
\end{figure}

In Fig.~\ref{fig:demographic_diff}, we compare each mobility indicator of individuals who fall in a particular socio-demographic category against the population of employed individuals. The results show significant differences between male and female individuals across all three mobility indicators, namely radius of gyration ($r_g$), time allocation entropy ($H$), and capacity ($C$). Unemployed women generally exhibit lower values of mobility exploration ($r_g$, and $C$) and diversity ($H$ and $C$).
Regarding individuals'\textit{Age}, differences in mobility behaviour are relatively smaller, with older individuals ($age \geq 55$) showing a more pronounced reduction in their characteristic geographical displacement ($r_g$) compared to other groups. 
\textit{Income} disparities reveal smaller differences in the radius of gyration ($r_g$), whereas richer individuals ($earnings \geq \$3333/month$) exhibit lower values in their time allocation entropy ($H$). Capacity ($C$), in contrast, is lower for those individuals reporting lower income values ($earnings \leq \$1250/month$). 
In terms of \textit{Race}, \textit{Asians} display lower values in all three mobility metrics, followed by \textit{Black or African American} individuals and then \textit{White} individuals. Specific ethnic groups (e.g., \textit{American Indian or Alaska Native}, \textit{Native Hawaiian or Other Pacific Islander} and \textit{Two or More Race Groups}) have been excluded from the analysis due to small sample sizes.
\textit{Educational} levels show fewer differences in mobility behaviour between groups, with no significant differences in radius of gyration ($r_g$). Lower values of time allocation entropy ($H$) and capacity ($C$) are observed in individuals with \textit{Bachelor degree or advanced degrees}.

We test the significance of the differences between demographic groups using Welch’s t-test~\cite{welch1947generalization} (see SI Tab. S4), considering the behavioural information from $30$ days after job-loss up to $100$ days (to remove the initial transitioning phase). 

To validate our understanding of job loss as an important life-course event that can shape individual mobility patterns, we conducted a comparative analysis between unemployed individuals and their employed counterparts from the same socio-demographic group. This comparison demonstrates that the reduced mobility behaviour occurring after a job loss is consistently present, although with different intensity, in all the studied population strata (see SI Tab. S5 for the statistics).

Taken together, these results substantiate the interpretation of socio-demographic characteristics as a factor to be taken into account when aiming to mitigate the effects of unemployment on individuals in mobility patterns. 
The differences observed in mobility indicators after a job loss are attributable to both the transition to joblessness and the inherent mobility tendencies within socio-demographic groups~\cite{lenormand2015influence,gauvin2020gender, gauvin2021socio,deng2021high}.

\section*{Discussion}
The availability of massive digital traces collected through mobile phones has become an important proxy for studying individual behaviour at population scales. The size and granularity of these datasets have revealed crucial insights into the regularities of human mobility and have exposed universal properties of human mobility patterns \cite{song2010limits, song2010modelling, simini2012universal, barbosa2018human, alessandretti2018evidence, alessandretti2020scales, schlapfer2021universal}. Interestingly, within the numerous mobility models for human mobility, the notion of ``opportunities'' consistently emerges as a key driver of individual movement patterns \cite{stouffer1940intervening, erlander1990gravity, gonzalez2008understanding, song2010modelling, simini2012universal,lucchini2019following, pappalardo2016human}. This notion suggests that individuals navigate physical space in pursuit of various kinds of opportunities spanning social, educational, and economic domains. 

In this perspective, understanding whether individuals transitioning to a state of unemployment can still access and benefit from the opportunities that their social and physical environments offer is of great social importance~\cite{paul2018individual,brand2015far,wanberg2012individual}.

To proxy social exposure and access to opportunities, we leverage an individual-level longitudinal dataset of fine-grained mobility behaviour alongside secondary demographic data~\cite{dingel2020many,MonthlyUI,CBG,NAICS}. 
We employ reweighting and rescaling techniques to address potential sample biases in the GPS data and to mitigate the effects of COVID-19 restrictions on mobility behaviour analysis.

The empirical evidence we present highlights that individuals facing unemployment significantly decrease their mobility, suggesting a reduction in their ability to explore and exploit available opportunities. 
This effect worsens over time, leading to a differentiation of the population into employed and unemployed subgroups with persistent behavioural differences.
In particular, the impact of job loss manifests differently across various socio-demographic groups, highlighting how some of these already vulnerable communities may be disproportionately affected~\cite{gauvin2020gender,gauvin2021socio,lenormand2015influence,wanberg2012individual, deng2021high}. 

In this context, our work underscores the significant influence of personal circumstances or life events, such as job loss, on established patterns of human mobility. 
These life-course events can drive individuals to transition through different states of human mobility regularities, adding a layer of complexity to the notion that mobility patterns can depend on the structure of the surrounding physical space~\cite{barbosa2021uncovering,thiemann2010structure,alessandretti2020scales} and the demographic attributes of individuals~\cite{barbosa2018human,barbosa2021uncovering,alessandretti2020scales}.

Furthermore, for an individual, the reduction in exploration patterns is not only a reflection of the immediate impact of job loss but potentially also signals a broader issue, leading to a decreased exposure to opportunities and an increased risk of social isolation after experiencing job loss~\cite{dieckhoff2015unemployed,pohlan2019unemployment,toole2015coupling,lucchini2021living}, reinforcing negative effects on an individual's well-being \cite{mckee2005psychological, warr1987work, warr1988unemployment, feather2012psychological}. 

Importantly, the progressive reduction in individual mobility and the associated decline in social participation~\cite{dieckhoff2015unemployed,pohlan2019unemployment,toole2015coupling,lucchini2021living}, could also undermine the potential effectiveness of intervention programs targeting the early stages of unemployment. 
By leveraging real-time data, our approach can facilitate targeted efforts during these initial phases, enabling more effective mitigation of the enduring and group-specific impacts of job loss~\cite{wanberg2012individual,brand2015far}.

\section*{Methods}
\subsection*{Stop locations}
The GPS location data was provided by Cuebiq, a location intelligence company that provided through their Cuebiq Data for Good COVID-19 Collaborative program, a dataset of privacy-enhanced GPS locations from users who opted-in to share the data anonymously for research purposes through a CCPA (California Consumer Privacy Act) compliant framework (see SI S1 for more details). To further preserve privacy, the data provider obfuscates users' precise home and work locations by transforming them to the centroid of the corresponding Census Block Group. 
We analyze a dataset that spans a period of 9 months, from January 2020 to September 2020 for seven US states including New York, Wyoming, Indiana, Idaho, Washington, North Dakota, and New Mexico.

We filter out all users with less than one month of data before declaring a national emergency (March 13, 2020) and less than four months after it. We also require users to have 5 hours per day covered by at least one GPS location. The resulting dataset includes more than 1 million anonymous, opted-in individuals. 
For all users, we extract their stop events with an algorithm based on Hariharan and Toyama~\cite{hariharan2004project}. We define a stop event as a temporal sequence of GPS coordinates in a radius $\Delta_s=65$ meters where a user stayed for at least $\Delta_t = 5$ minutes. 
For each user, we then define their stop locations as the stop events that can be considered as part of the same place using the DBSCAN algorithm~\cite{ester1996density}. With DBSCAN, we group points within a distance of $\epsilon = \Delta_s - 5$ meters to form a cluster with at least $\text{minPoints}=1$ stop event. For a more detailed explanation of the GPS data processing please refer to Lucchini et al.~\cite{lucchini2021living}.

\subsection*{Residential and Workplace detection} 
We determined the most likely residential and workplace areas for each user by calculating these areas multiple times over a moving rolling window of 28 days. 
We then aggregate for each day $t$, for each user $u$, and stop $s_u(t)$, the amount of time spent in a window $[t-14, t+14]$ distinguishing between:
\begin{enumerate}
    \item \textbf{Residential time:} The amount of time between 8 pm and 4 am spent by the user $u$ at stop $s_u$. Unlike previous studies such as TimeGeo~\cite{jiang2016timegeo}, we did not assume the entire weekend as residential time since the US Bureau of Labor Statistics recently estimated that around 34\% of employed people work in the weekend~\cite{bureau}.
    \item \textbf{Workplace time:} On weekdays, the amount of time between 9 am and 5 pm spent by the user $u$ at stop $s_u$. We chose these working hours because they represent the most common working time in the US~\cite{officeHoursUS}. Additionally, we assumed that a potential workplace stay should last at least 30 minutes and occur five times a week. These assumptions were similar to those made in previous studies~\cite{jiang2016timegeo}. 
\end{enumerate}
We detect for each user $u$ their \textbf{residential location} as the stop location $s_u(t)$ with the largest \emph{Residential time} during the period that goes from January 3rd to March 7th (before the pandemic).
To detect changes in the \textbf{workplace location} for a particular user $u$, we label a stop $s_u(t)$ as workplace location if this stop is not a residential location (to avoid tracking people already working from home) and it has the largest \emph{Workplace time} in the observed 28 days window. 
To protect users' privacy, the residential and workplace locations were blurred and associated with the corresponding Census Block Groups.

\subsection*{Job assignment}\label{sec:job_assignment}
The location data does not have direct information about the users’ jobs. To the extracted residential and workplace locations, we associate a Geographic Identifier (GEOID), which is a numeric code that uniquely identifies an administrative geographic area in the US Census data. 
To be able to assign a job to each user, we match the residential and workplace GEOIDs to the GEOIDs of the Longitudinal Employer-Household Dynamics Origin-Destination Employment Statistics (LODES) datasets \cite{LODES}. Given the residential/workplace locations, these datasets provide statistics about the number of jobs in each sector as defined by the North American Industry Classification System (NAICS) \cite{NAICS}.

The information of the LODES datasets is organized into i) Residence Area Characteristics (RAC); ii) Workplace Area Characteristics (WAC); iii) Origin-Destination (OD). 
RAC and WAC datasets provide job statistics according to the residential and workplace census block groups respectively. The OD dataset provides job statistics considering both residential and workplace census blocks (for further details on the LODES datasets refer to SI S1).
From these three datasets, we then compute the probabilities of working in a particular NAICS sector by normalizing the number of jobs in each NAICS sector by the total number of jobs.

Finally, for each individual and the combination of their residential and workplace GEOIDs, we assign an industrial sector in probability. The probability of working in a specific sector, $i$, is computed for each user $u$ with a home location and a work location, as the joint probability of independently working in that specific sector, given that $u$ resides in $r$ and works in $w$ (their home and work locations respectively):
\begin{equation*}
    P(i, u | r, w) = P_{RAC}(i, u | r, w) \cdot P_{WAC}(i, u | r, w) \cdot P_{OD}(i, u | r, w)
\end{equation*} 
where $r$ is the residential GEOID and $w$ is the workplace GEOID as provided in the LODES data.

Bootstrapping is applied for a more robust NAICS assignment to individuals. Specifically, for each individual, we sample $10$ times from the corresponding NAICS probability distribution and retain all the sampled NAICS as independent realisations of a GEOID-representative population.

\subsection*{Employment status inference}\label{sec:unemployment_status}
To infer the employment status of a bootstrapped user, we leverage information on the reduction in i) workplace visits and ii) time spent at work. 

\subsubsection*{Who is at risk of unemployment?}
    We define an individual to be at risk of losing their job if they never visited their work location.
    Since we are interested in studying the impact of job loss at the individual level, we restrict our analysis to individuals who were employed before the pandemic declaration. Using the pre-pandemic period as a baseline period makes it possible to investigate the shock induced by the pandemic on the job market. Specifically, we retain only users who have been working at least 5 days during the baseline period ($t=b$), namely the period before the pandemic (January 3rd - March 7th, 2020). 
    We then compute the reduction of workplace visits and identify as ``at risk'' those who at a specific time window $t$ didn't visit their workplace. 
    To identify the population at risk of unemployment over time, we used a time window of $28$ days with a daily shift. Thus, at each time $t$, we define an individual to be at risk of unemployment, $r_u(t)$, if they didn't visit their work location in the entire time window:
    \begin{equation*}
    \begin{cases}
        r_u(t) = 1 & \text{if }v_u(t) = 0\\
        r_u(t) = 0 & \text{otherwise}
    \end{cases}
    \end{equation*}
    with $v_u(t)$ representing the number of visits a user $u$ made to their work location within the time window $t$.

\subsubsection*{Who is working remotely?}
    Under the assumption that individuals at risk of unemployment ($r_u(t) = 1$) could be working remotely, we assign to each individual the likelihood of working remotely based both on their personal and other workers' (in the same NAICS) working behaviour changes. Individual working behaviour change is measured in terms of the reduction in the time spent at work with respect to the baseline period:
    \begin{equation*}
        rt_u(t) = 1 -\frac{w_u(t)}{w_u(b)}.
    \end{equation*}
    Here $w_u(t)$ is the time the user $u$ spent at the workplace in the time window $t$, and $w_u(b)$ is the median of the time spent at work (within windows of the same size as $t$) during the baseline period (January 3rd - March 7th, 2020).

    By additionally adjusting for how much the entire sector $s$ is working remotely during a specific window compared the estimated maximum amount of time that can be worked remotely~\cite{dingel2020many}, we can write the probability of being unemployed as:
    \begin{equation*}
        P_s(t) = 1-\min\Bigg(\frac{\tilde{R}_s(t)}{\tilde{r}_{s}(t)}, 1\Bigg)
    \end{equation*}
    where $\tilde{r}_{s}(t)$ represents the weighted fraction of remaining work time that could be performed remotely by those individuals who stopped visiting their work location, and $\tilde{R}_s(t)$ represent the fraction of work that could be performed remotely adjusted by potential changes in remote working behaviour among those individuals who didn't interrupt visiting their workplace (for additional details see SI S3A).

    Intuitively, by estimating at the sector level the reduction in the time spent at work by users who are still visiting their work location ($r_u(t) < 1$), we measure how much ``remote work'' is already performed by individuals who are still visiting their workplace. The remaining part of the remote-workable time (if any), $\tilde{R}_s(t)$, is used to uniformly distribute the probability of being unemployed among individuals who stopped visiting their workplace (for further details see SI S3A).

\subsection*{Mobility metrics}
To track the changes in mobility of employed and unemployed individuals, we measure within-individual variations by comparing mobility behaviours to a baseline period preceding the pandemic (February 1st, 2020 - March 7th, 2020). The mobility metrics we used in our analysis offer a comprehensive picture of mobility behaviour including individuals' characteristic displacement and the complexity of individuals' exploratory behaviour. To track the changes over time, we computed these metrics over a moving window of 28 days with a 1-day shift. 

\subsubsection*{Time Allocation Entropy}\label{sec:entropy}
We introduce the time allocation entropy, which measures the distribution of time allocation in each visited location by an individual, as a measure of exploratory behaviour:
 $H(i)= -\frac{\sum_{j=1}^k p_{ij} \log(p_{ij})}{\log k}.$
Here, $k$ is the total number of unique visited locations of an individual $i$, $p_{ij} = \frac{V_{ij}}{\sum_{j=1}^k V_{ij}}$ and $V_{ij}$ is the total time individual $i$ spends in location $j$ (weighted by time spent).

\subsubsection*{Radius of gyration}\label{sec:radius_gyration}
To measure the characteristic geographical displacement of individuals, we use the well-known radius of gyration \cite{gonzalez2008understanding, pappalardo2015returners} defined as:
$r_g = \sqrt{\frac{1}{N} \sum_{l \in L} n_l|\mathbf{r_l} - \mathbf{r_{cm}}|^2}$, 
where $N$ is the total time spent by a particular individual to all their visited locations; $n_l$ is the time spent to location $l$; $L$ is the set of stop locations within a time window; $\mathbf{r_l}$ is a two-dimensional vector representing the location’s GPS position recorded as latitude and longitude; and $\mathbf{r_{cm}}$ is the centre of mass of the trajectories, defined as $\mathbf{r_{cm}} = \frac{1}{N} \sum_{l=1}^{N} \mathbf{r_l}$.

\subsubsection*{Capacity}
We capture and track the number of an individual's familiar locations following the definition of Alessandretti et al. \cite{alessandretti2018evidence}. For each individual, we compute the location capacity $C$ in each time window, normalized by the mean capacity of all users during the baseline period before the pandemic (January 3rd - March 7th, 2020). 

Together with the capacity $C$, we also computed the number of locations added to ($A$) and deleted from ($D$) the set of familiar locations within a specific time interval and the previous time interval~\cite{alessandretti2018evidence}.

\section*{Data and code availability}
The data supporting the findings of this study are accessible through Cuebiq’s Data for Good initiative. For details on how to request access, including conditions and limitations, please visit: \url{https://www.cuebiq.com/about/data-for-good/}.

\noindent
Replication code is available on GitHub at \url{https://github.com/scentellegher/ImpactJobLoss/}.



\section*{Acknowledgements}
The authors would like to thank Cuebiq that kindly provided us with the mobility dataset for this research through their Data for Good program.
L.L. thanks G.K. for the insightful discussions and his support during the entire project development.
L.L. has been supported by the ERC project ``IMMUNE'' (Grant agreement ID: 101003183). L.L. acknowledges the support from the ``Fondazione Romeo ed Enrica Invernizzi'' for the research activities of the 'Covid Crisis Lab' at Bocconi University.
S.C. and B.L. have been supported by the PNRR ICSC National Research Centre for High Performance Computing, Big Data and Quantum Computing (CN00000013), under the NRRP MUR program funded by the NextGenerationEU.

\section*{Author contributions statement}
L.L., S.C., M.D.N. conceived the original idea and planned the experiments. S.C., L.L. and M.D.N. pre-processed the mobility data. S.C., L.L. and M.T. carried out the experiments and made the Figures. L.L., S.C. and M.D.N. contributed to the interpretation of the results. L.L. and S.C. wrote the manuscript. S.C., M.D.N., M.T., B.L., and L.L. provided critical feedback, helped shape the manuscript and substantively revised it.

\clearpage
\bibliographystyle{unsrt}
\bibliography{sample}

\begin{thebibliography}{10}

\bibitem{bls_unemployment}
{U.S. Bureau of Labor Statistics}.
\newblock {How the Government Measures Unemployment}.
\newblock \url{https://www.bls.gov/cps/cps_htgm.htm}, 2023.

\bibitem{krueger2017evolution}
Alan~B Krueger, Alexandre Mas, and Xiaotong Niu.
\newblock The evolution of rotation group bias: Will the real unemployment rate
  please stand up?
\newblock {\em Review of Economics and Statistics}, 99(2):258--264, 2017.

\bibitem{ons2023labour}
{Office for National Statistics (ONS)}.
\newblock {Labour Force Survey performance and quality monitoring report: April
  to June 2023}.
\newblock \url{
  https://www.ons.gov.uk/employmentandlabourmarket/peopleinwork/employmentandemployeetypes/methodologies/labourforcesurveyperformanceandqualitymonitoringreport},
  2023.

\bibitem{international2015world}
International~Labour Office.
\newblock {\em World employment and social outlook: Trends 2015}.
\newblock International Labour Organization Geneva, 2015.

\bibitem{dewan2022world}
Sabina Dewan, Ekkehard Ernst, and Souleima Achkar~Hilal.
\newblock World employment and social outlook: trends 2022, 2022.

\bibitem{gonzalez2008understanding}
Marta~C Gonzalez, Cesar~A Hidalgo, and Albert-Laszlo Barabasi.
\newblock Understanding individual human mobility patterns.
\newblock {\em nature}, 453(7196):779--782, 2008.

\bibitem{pappalardo2015returners}
Luca Pappalardo, Filippo Simini, Salvatore Rinzivillo, Dino Pedreschi, Fosca
  Giannotti, and Albert-L{\'a}szl{\'o} Barab{\'a}si.
\newblock Returners and explorers dichotomy in human mobility.
\newblock {\em Nature communications}, 6(1):8166, 2015.

\bibitem{lucchini2019following}
Lorenzo Lucchini, Sara Tonelli, and Bruno Lepri.
\newblock Following the footsteps of giants: modeling the mobility of
  historically notable individuals using wikipedia.
\newblock {\em EPJ Data Science}, 8(1):36, 2019.

\bibitem{alessandretti2020scales}
Laura Alessandretti, Ulf Aslak, and Sune Lehmann.
\newblock The scales of human mobility.
\newblock {\em Nature}, 587(7834):402--407, 2020.

\bibitem{singh2015money}
Vivek~Kumar Singh, Burcin Bozkaya, and Alex Pentland.
\newblock Money walks: implicit mobility behavior and financial well-being.
\newblock {\em PloS one}, 10(8):e0136628, 2015.

\bibitem{tovanich2021inferring}
Natkamon Tovanich, Simone Centellegher, Nac{\'e}ra~Bennacer Seghouani, Joe
  Gladstone, Sandra Matz, and Bruno Lepri.
\newblock Inferring psychological traits from spending categories and dynamic
  consumption patterns.
\newblock {\em EPJ Data Science}, 10(1):24, 2021.

\bibitem{dong2017social}
Xiaowen Dong, Yoshihiko Suhara, Bur{\c{c}}in Bozkaya, Vivek~K Singh, Bruno
  Lepri, and Alex~‘Sandy’ Pentland.
\newblock Social bridges in urban purchase behavior.
\newblock {\em ACM Transactions on Intelligent Systems and Technology (TIST)},
  9(3):1--29, 2017.

\bibitem{matz2016money}
Sandra~C Matz, Joe~J Gladstone, and David Stillwell.
\newblock Money buys happiness when spending fits our personality.
\newblock {\em Psychological science}, 27(5):715--725, 2016.

\bibitem{lucchini2022reddit}
Lorenzo Lucchini, Luca~Maria Aiello, Laura Alessandretti, Gianmarco
  De~Francisci~Morales, Michele Starnini, and Andrea Baronchelli.
\newblock From reddit to wall street: The role of committed minorities in
  financial collective action.
\newblock {\em Royal Society Open Science}, 9(4):211488, 2022.

\bibitem{sobolevsky2016cities}
Stanislav Sobolevsky, Izabela Sitko, Remi Tachet~des Combes, Bartosz Hawelka,
  Juan Murillo~Arias, and Carlo Ratti.
\newblock Cities through the prism of people’s spending behavior.
\newblock {\em PloS one}, 11(2):e0146291, 2016.

\bibitem{wang2018urban}
Qi~Wang, Nolan~Edward Phillips, Mario~L Small, and Robert~J Sampson.
\newblock Urban mobility and neighborhood isolation in america’s 50 largest
  cities.
\newblock {\em Proceedings of the National Academy of Sciences},
  115(30):7735--7740, 2018.

\bibitem{chetty2022sociala}
Raj Chetty, Matthew~O Jackson, Theresa Kuchler, Johannes Stroebel, Nathaniel
  Hendren, Robert~B Fluegge, Sara Gong, Federico Gonzalez, Armelle Grondin,
  Matthew Jacob, et~al.
\newblock Social capital i: measurement and associations with economic
  mobility.
\newblock {\em Nature}, 608(7921):108--121, 2022.

\bibitem{chetty2022socialb}
Raj Chetty, Matthew~O Jackson, Theresa Kuchler, Johannes Stroebel, Nathaniel
  Hendren, Robert~B Fluegge, Sara Gong, Federico Gonzalez, Armelle Grondin,
  Matthew Jacob, et~al.
\newblock Social capital ii: determinants of economic connectedness.
\newblock {\em Nature}, 608(7921):122--134, 2022.

\bibitem{moro2021mobility}
Esteban Moro, Dan Calacci, Xiaowen Dong, and Alex Pentland.
\newblock Mobility patterns are associated with experienced income segregation
  in large us cities.
\newblock {\em Nature communications}, 12(1):4633, 2021.

\bibitem{yabe2023behavioral}
Takahiro Yabe, Bernardo Garc{\'\i}a~Bulle Bueno, Xiaowen Dong, Alex Pentland,
  and Esteban Moro.
\newblock Behavioral changes during the covid-19 pandemic decreased income
  diversity of urban encounters.
\newblock {\em Nature Communications}, 14(1):2310, 2023.

\bibitem{wang2016crime}
Hongjian Wang, Daniel Kifer, Corina Graif, and Zhenhui Li.
\newblock Crime rate inference with big data.
\newblock In {\em Proceedings of the 22nd ACM SIGKDD international conference
  on knowledge discovery and data mining}, pages 635--644, 2016.

\bibitem{song2019crime}
Guangwen Song, Wim Bernasco, Lin Liu, Luzi Xiao, Suhong Zhou, and Weiwei Liao.
\newblock Crime feeds on legal activities: Daily mobility flows help to explain
  thieves’ target location choices.
\newblock {\em Journal of Quantitative Criminology}, 35:831--854, 2019.

\bibitem{luca2023crime}
Massimiliano Luca, Gian~Maria Campedelli, Simone Centellegher, Michele Tizzoni,
  and Bruno Lepri.
\newblock Crime, inequality and public health: A survey of emerging trends in
  urban data science.
\newblock {\em Frontiers in Big Data}, 6:50, 2023.

\bibitem{wesolowski2012quantifying}
Amy Wesolowski, Nathan Eagle, Andrew~J Tatem, David~L Smith, Abdisalan~M Noor,
  Robert~W Snow, and Caroline~O Buckee.
\newblock Quantifying the impact of human mobility on malaria.
\newblock {\em Science}, 338(6104):267--270, 2012.

\bibitem{oliver2020mobile}
Nuria Oliver, Bruno Lepri, Harald Sterly, Renaud Lambiotte, S{\'e}bastien
  Deletaille, Marco De~Nadai, Emmanuel Letouz{\'e}, Albert~Ali Salah, Richard
  Benjamins, Ciro Cattuto, et~al.
\newblock Mobile phone data for informing public health actions across the
  covid-19 pandemic life cycle.
\newblock {\em Science advances}, 6(23):eabc0764, 2020.

\bibitem{kraemer2020effect}
Moritz~UG Kraemer, Chia-Hung Yang, Bernardo Gutierrez, Chieh-Hsi Wu, Brennan
  Klein, David~M Pigott, Open COVID-19 Data~Working Group†, Louis Du~Plessis,
  Nuno~R Faria, Ruoran Li, et~al.
\newblock The effect of human mobility and control measures on the covid-19
  epidemic in china.
\newblock {\em Science}, 368(6490):493--497, 2020.

\bibitem{aleta2020modelling}
Alberto Aleta, David Martin-Corral, Ana Pastore~y Piontti, Marco Ajelli, Maria
  Litvinova, Matteo Chinazzi, Natalie~E Dean, M~Elizabeth Halloran, Ira~M
  Longini~Jr, Stefano Merler, et~al.
\newblock Modelling the impact of testing, contact tracing and household
  quarantine on second waves of covid-19.
\newblock {\em Nature Human Behaviour}, 4(9):964--971, 2020.

\bibitem{moriwaki2020nowcasting}
Daisuke Moriwaki.
\newblock Nowcasting unemployment rates with smartphone gps data.
\newblock In {\em Multiple-Aspect Analysis of Semantic Trajectories: First
  International Workshop, MASTER 2019, Held in Conjunction with ECML-PKDD 2019,
  W{\"u}rzburg, Germany, September 16, 2019, Proceedings 1}, pages 21--33.
  Springer, 2020.

\bibitem{sundsoy2016estimating}
P{\aa}l Sunds{\o}y, Johannes Bjelland, Bj{\o}rn-Atle Reme, Eaman Jahani, Erik
  Wetter, and Linus Bengtsson.
\newblock Estimating individual employment status using mobile phone network
  data.
\newblock {\em arXiv preprint arXiv:1612.03870}, 2016.

\bibitem{toole2015tracking}
Jameson~L Toole, Yu-Ru Lin, Erich Muehlegger, Daniel Shoag, Marta~C
  Gonz{\'a}lez, and David Lazer.
\newblock Tracking employment shocks using mobile phone data.
\newblock {\em Journal of The Royal Society Interface}, 12(107):20150185, 2015.

\bibitem{almaatouq2016mobile}
Abdullah Almaatouq, Francisco Prieto-Castrillo, and Alex Pentland.
\newblock Mobile communication signatures of unemployment.
\newblock In {\em Social Informatics: 8th International Conference, SocInfo
  2016, Bellevue, WA, USA, November 11-14, 2016, Proceedings, Part I 8}, pages
  407--418. Springer, 2016.

\bibitem{barbosa2021uncovering}
Hugo Barbosa, Surendra Hazarie, Brian Dickinson, Aleix Bassolas, Adam Frank,
  Henry Kautz, Adam Sadilek, Jos{\'e}~J Ramasco, and Gourab Ghoshal.
\newblock Uncovering the socioeconomic facets of human mobility.
\newblock {\em Scientific reports}, 11(1):8616, 2021.

\bibitem{noulas2012tale}
Anastasios Noulas, Salvatore Scellato, Renaud Lambiotte, Massimiliano Pontil,
  and Cecilia Mascolo.
\newblock A tale of many cities: universal patterns in human urban mobility.
\newblock {\em PloS one}, 7(5):e37027, 2012.

\bibitem{alessandretti2018evidence}
Laura Alessandretti, Piotr Sapiezynski, Vedran Sekara, Sune Lehmann, and Andrea
  Baronchelli.
\newblock Evidence for a conserved quantity in human mobility.
\newblock {\em Nature human behaviour}, 2(7):485--491, 2018.

\bibitem{schlapfer2021universal}
Markus Schl{\"a}pfer, Lei Dong, Kevin O’Keeffe, Paolo Santi, Michael Szell,
  Hadrien Salat, Samuel Anklesaria, Mohammad Vazifeh, Carlo Ratti, and
  Geoffrey~B West.
\newblock The universal visitation law of human mobility.
\newblock {\em Nature}, 593(7860):522--527, 2021.

\bibitem{barbosa2018human}
Hugo Barbosa, Marc Barthelemy, Gourab Ghoshal, Charlotte~R James, Maxime
  Lenormand, Thomas Louail, Ronaldo Menezes, Jos{\'e}~J Ramasco, Filippo
  Simini, and Marcello Tomasini.
\newblock Human mobility: Models and applications.
\newblock {\em Physics Reports}, 734:1--74, 2018.

\bibitem{LODES}
Center for Economic~Studies US~Census~Bureau.
\newblock Data - longitudinal employer-household dynamics.
\newblock Accessed on 2024-02-02.

\bibitem{NAICS}
North american industry classification system (naics) u.s. census bureau.
\newblock Accessed on 2024-02-02.

\bibitem{CBG}
US~Census Bureau.
\newblock Glossary.
\newblock Accessed on 2024-02-02.

\bibitem{dingel2020many}
Jonathan~I Dingel and Brent Neiman.
\newblock How many jobs can be done at home?
\newblock {\em Journal of Public Economics}, 189:104235, 2020.

\bibitem{lucchini2023socioeconomic}
Lorenzo Lucchini, Ollin Langle-Chimal, Lorenzo Candeago, Lucio Melito, Alex
  Chunet, Aleister Montfort, Bruno Lepri, Nancy Lozano-Gracia, and Samuel~P
  Fraiberger.
\newblock Socioeconomic disparities in mobility behavior during the covid-19
  pandemic in developing countries.
\newblock {\em arXiv preprint arXiv:2305.06888}, 2023.

\bibitem{song2010limits}
Chaoming Song, Zehui Qu, Nicholas Blumm, and Albert-L{\'a}szl{\'o}
  Barab{\'a}si.
\newblock Limits of predictability in human mobility.
\newblock {\em Science}, 327(5968):1018--1021, 2010.

\bibitem{llorente2015social}
Alejandro Llorente, Manuel Garcia-Herranz, Manuel Cebrian, and Esteban Moro.
\newblock Social media fingerprints of unemployment.
\newblock {\em PloS one}, 10(5):e0128692, 2015.

\bibitem{welch1947generalization}
Bernard~L Welch.
\newblock The generalization of ‘student's’problem when several different
  population varlances are involved.
\newblock {\em Biometrika}, 34(1-2):28--35, 1947.

\bibitem{lenormand2015influence}
Maxime Lenormand, Thomas Louail, Oliva~G Cant{\'u}-Ros, Miguel Picornell,
  Ricardo Herranz, Juan~Murillo Arias, Marc Barthelemy, Maxi~San Miguel, and
  Jos{\'e}~J Ramasco.
\newblock Influence of sociodemographic characteristics on human mobility.
\newblock {\em Scientific reports}, 5(1):10075, 2015.

\bibitem{gauvin2020gender}
Laetitia Gauvin, Michele Tizzoni, Simone Piaggesi, Andrew Young, Natalia Adler,
  Stefaan Verhulst, Leo Ferres, and Ciro Cattuto.
\newblock Gender gaps in urban mobility.
\newblock {\em Humanities and Social Sciences Communications}, 7(1):1--13,
  2020.

\bibitem{gauvin2021socio}
Laetitia Gauvin, Paolo Bajardi, Emanuele Pepe, Brennan Lake, Filippo Privitera,
  and Michele Tizzoni.
\newblock Socio-economic determinants of mobility responses during the first
  wave of covid-19 in italy: from provinces to neighbourhoods.
\newblock {\em Journal of The Royal Society Interface}, 18(181):20210092, 2021.

\bibitem{deng2021high}
Hengfang Deng, Daniel~P Aldrich, Michael~M Danziger, Jianxi Gao, Nolan~E
  Phillips, Sean~P Cornelius, and Qi~Ryan Wang.
\newblock High-resolution human mobility data reveal race and wealth
  disparities in disaster evacuation patterns.
\newblock {\em Humanities and Social Sciences Communications}, 8(1):1--8, 2021.

\bibitem{song2010modelling}
Chaoming Song, Tal Koren, Pu~Wang, and Albert-L{\'a}szl{\'o} Barab{\'a}si.
\newblock Modelling the scaling properties of human mobility.
\newblock {\em Nature physics}, 6(10):818--823, 2010.

\bibitem{simini2012universal}
Filippo Simini, Marta~C Gonz{\'a}lez, Amos Maritan, and Albert-L{\'a}szl{\'o}
  Barab{\'a}si.
\newblock A universal model for mobility and migration patterns.
\newblock {\em Nature}, 484(7392):96--100, 2012.

\bibitem{stouffer1940intervening}
Samuel~A. Stouffer.
\newblock Intervening opportunities: A theory relating mobility and distance.
\newblock {\em American Sociological Review}, 5(6):845--867, 1940.

\bibitem{erlander1990gravity}
Sven Erlander and Neil~F Stewart.
\newblock {\em The gravity model in transportation analysis: theory and
  extensions}, volume~3.
\newblock Vsp, 1990.

\bibitem{pappalardo2016human}
Luca Pappalardo, Salvatore Rinzivillo, and Filippo Simini.
\newblock Human mobility modelling: exploration and preferential return meet
  the gravity model.
\newblock {\em Procedia Computer Science}, 83:934--939, 2016.

\bibitem{paul2018individual}
Karsten~I Paul, Alice Hassel, and Klaus Moser.
\newblock Individual consequences of job loss and unemployment.
\newblock {\em Oxford handbook of job loss and job search}, pages 57--85, 2018.

\bibitem{brand2015far}
Jennie~E Brand.
\newblock The far-reaching impact of job loss and unemployment.
\newblock {\em Annual review of sociology}, 41:359--375, 2015.

\bibitem{wanberg2012individual}
Connie~R Wanberg.
\newblock The individual experience of unemployment.
\newblock {\em Annual review of psychology}, 63:369--396, 2012.

\bibitem{MonthlyUI}
Monthly program and financial data, employment \& training administration (eta)
  - u.s. department of labor.
\newblock Accessed on 2024-01-31.

\bibitem{thiemann2010structure}
Christian Thiemann, Fabian Theis, Daniel Grady, Rafael Brune, and Dirk
  Brockmann.
\newblock The structure of borders in a small world.
\newblock {\em PloS one}, 5(11):e15422, 2010.

\bibitem{dieckhoff2015unemployed}
Martina Dieckhoff and Vanessa Gash.
\newblock Unemployed and alone? unemployment and social participation in
  europe.
\newblock {\em International Journal of Sociology and Social Policy},
  35(1/2):67--90, 2015.

\bibitem{pohlan2019unemployment}
Laura Pohlan.
\newblock Unemployment and social exclusion.
\newblock {\em Journal of Economic Behavior \& Organization}, 164:273--299,
  2019.

\bibitem{toole2015coupling}
Jameson~L Toole, Carlos Herrera-Yaq{\"u}e, Christian~M Schneider, and Marta~C
  Gonz{\'a}lez.
\newblock Coupling human mobility and social ties.
\newblock {\em Journal of The Royal Society Interface}, 12(105):20141128, 2015.

\bibitem{lucchini2021living}
Lorenzo Lucchini, Simone Centellegher, Luca Pappalardo, Riccardo Gallotti,
  Filippo Privitera, Bruno Lepri, and Marco De~Nadai.
\newblock Living in a pandemic: changes in mobility routines, social activity
  and adherence to covid-19 protective measures.
\newblock {\em Scientific reports}, 11(1):24452, 2021.

\bibitem{mckee2005psychological}
Frances McKee-Ryan, Zhaoli Song, Connie~R Wanberg, and Angelo~J Kinicki.
\newblock Psychological and physical well-being during unemployment: a
  meta-analytic study.
\newblock {\em Journal of applied psychology}, 90(1):53, 2005.

\bibitem{warr1987work}
Peter Warr.
\newblock {\em Work, unemployment, and mental health.}
\newblock Oxford University Press, 1987.

\bibitem{warr1988unemployment}
Peter Warr, Paul Jackson, and Michael Banks.
\newblock Unemployment and mental health: Some british studies.
\newblock {\em Journal of social issues}, 44(4):47--68, 1988.

\bibitem{feather2012psychological}
Norman~T Feather.
\newblock {\em The psychological impact of unemployment}.
\newblock Springer Science \& Business Media, 2012.

\bibitem{hariharan2004project}
Ramaswamy Hariharan and Kentaro Toyama.
\newblock Project lachesis: Parsing and modeling location histories.
\newblock In Max~J. Egenhofer, Christian Freksa, and Harvey~J. Miller, editors,
  {\em Geographic Information Science}, pages 106--124, Berlin, Heidelberg,
  2004. Springer Berlin Heidelberg.

\bibitem{ester1996density}
Martin Ester, Hans-Peter Kriegel, J{\"o}rg Sander, Xiaowei Xu, et~al.
\newblock A density-based algorithm for discovering clusters in large spatial
  databases with noise.
\newblock In {\em kdd}, volume 96, 34, pages 226--231, 1996.

\bibitem{jiang2016timegeo}
Shan Jiang, Yingxiang Yang, Siddharth Gupta, Daniele Veneziano, Shounak
  Athavale, and Marta~C Gonz{\'a}lez.
\newblock The timegeo modeling framework for urban mobility without travel
  surveys.
\newblock {\em Proceedings of the National Academy of Sciences},
  113(37):E5370--E5378, 2016.

\bibitem{bureau}
{Bureau of Labor Statistics, American Time Use Survey}.
\newblock {Percent of population who worked on weekdays and weekend days}.
\newblock \url{ https://www.bls.gov/tus/charts/chart11.pdf}, 2015.

\bibitem{officeHoursUS}
Wikipedia.
\newblock Business hours - wikipedia.
\newblock Version access: 2023-12-24.

\end{thebibliography}



\end{document}


\title{Supplementary Information of: Job loss disrupts individuals' mobility and their exploratory patterns}

\author{Simone Centellegher\textsuperscript{*}}
\affiliation{Fondazione Bruno Kessler (FBK), Trento, Italy}

\author{Marco De Nadai}
\affiliation{Fondazione Bruno Kessler (FBK), Trento, Italy}

\author{Marco Tonin}
\affiliation{Fondazione Bruno Kessler (FBK), Trento, Italy}
\affiliation{Department of Sociology and Social Research, University of Trento, Trento, Italy}

\author{Bruno Lepri}
\affiliation{Fondazione Bruno Kessler (FBK), Trento, Italy}

\author{Lorenzo Lucchini\textsuperscript{*}}
\affiliation{Fondazione Bruno Kessler (FBK), Trento, Italy}
\affiliation{Centre for Social Dynamics and Public Policy, Bocconi University, Milan 20100, Italy}
\affiliation{Institute for Data Science and Analytics, Bocconi University, Milan 20100, Italy}

\thanks{These authors contributed equally to this work.}

\maketitle
\tableofcontents

\newpage
\section{Datasets}
\label{SI:sec:datasets}
\subsection{GPS location data}
The location data is provided by Cuebiq Inc., a location intelligence and measurement company. The dataset was shared within the Cuebiq Data for Good program, which provides access to de-identified and anonymized mobility data for academic and research purposes. 

The location data provided consists of privacy-enhanced GPS locations for research purposes, from January 2020 to September 2020, and includes only users who have opted-in to share their data anonymously. The data is General Data Protection Regulation (GDPR) and California Consumer Privacy Act (CCPA) compliant. Furthermore, to increase and preserve users' privacy, Cuebiq obfuscates home and work locations to the Census Block Group level. 
The data is collected through the Cuebiq Software Development Kit (SDK) which collects user locations through GPS and Wi-Fi signals in Android and iOS devices.

\subsection{Longitudinal Employer-Household Dynamics (LEHD)}
The Longitudinal Employer-Household Dynamics (LEHD) program of the US Census Bureau produces public-use information about employers and employees by combining federal, state, and Census Bureau data. Socio-economic information is used by state and local authorities to make informed decisions for their economies. We use three different surveys provided by this program in order to assign in probability industrial sectors to each individual analysed, according to the North American Industry Classification System (NAICS). NAICS represents the standard classification of businesses used by Federal Agencies in the
United States for the purposes of collecting and analysing statistical data related to the US business economy.
This study uses the LEHD Origin-Destination Employment Statistics (LODES) datasets, which include three different types of data: Origin-Destination (OD), Residence Area Characteristics (RAC), and Workplace Area Characteristics (WAC). The data was collected at the census block geographic level in 2018. LODES statistics contain data about geographic employment patterns by workplace and residential locations, and they include the age of the worker, earnings, industry, sex, race, ethnicity, and education. Therefore, LODES provide geographical statistics about employers and employees and can be used to answer questions about spatial, economic, and demographic issues related to workplaces and home-to-work flows.
The Residence Area Characteristics (RAC) and the Workplace Area Characteristics (WAC) datasets provide statistics about the total number of jobs, the total number of jobs in each NAICS sector, and also according to age, earnings of the workers, race, ethnicity, education, and sex. The difference between the RAC and the WAC datasets is that the first
computes these statistics according to the home census block, while in the WAC dataset jobs are totalled by the work census block.
The Origin-Destination (OD) dataset provides information considering both home and work census blocks. It includes the home and work GEOIDs and the same statistics of the WAC and the RAC datasets. However, different from the WAC and the RAC datasets, the OD
dataset provides the total number of jobs according to three macro-sectors, which are: Goods Producing industry sectors, Trade, Transportation, and Utilities industry sectors, and All Other Services industry sectors. These macro-sectors can be matched with the 20 NAICS sectors to combine the WAC, the RAC, and the OD datasets.

\subsection{Unemployment Insurance (UI) claims}
Unemployment Insurance (UI) programs are organised at the state level and they have the aim to provide assistance to jobless people who are looking for work. UI data provides the number of submitted claims divided by industry sectors (NAICS) and allows tracking of employment changes at the state level. 
This study uses the dataset ETA 203 - Characteristics of the Insured Unemployed of the UI program\footnote{https://oui.doleta.gov/unemploy/DataDownloads.asp}. It provides information about Unemployed Insurance claimants for each state for each month. It describes how the population of claimants varies over time and it reports characteristics about sex, race/ethnicity, age, industry, occupation, etc.

\subsection{Local Area Unemployment Statistics (LAUS)}
The Local Area Unemployment Statistics (LAUS) program of the Bureau of Labor Statistics (BLS)\footnote{https://www.bls.gov/lau/} provides monthly estimates of total employment and unemployment across different geographic levels (states, metropolitan areas, counties, etc). The employment and unemployment data are estimated by combining current and historical data from the Current Population Survey (CPS), the Current Employment Statistics (CES), and state Unemployment Insurance (UI) data. An estimation process done by the LAUS program produces the official unemployment rate.

\subsection{Employment data from the Bureau of Labor Statistics (BLS)}
The Bureau of Labor Statistics (BLS) is a unit of the Department of Labor of the United States that measures several economic aspects, such as labour market activity, working conditions, price changes, and productivity in the US economy. 
In this work, we incorporate state-level employment information obtained from the Bureau of Labor Statistics (BLS) through the Quarterly Census of Employment and Wages (QCEW) program. The program provides employment and wage information reported by employers with a coverage of more than 95 percent of US jobs. The data is available at the county, MSA, state, and national levels divided by NAICS sectors\footnote{https://www.bls.gov/cew/}.

\subsection{Remote Workability}
To include in our methodology for inferring the unemployment status of an individual the information about the teleworkability of a job, we leverage the work done by Dingel et al.~\cite{dingel2020many}. In their work, the authors developed a methodology and provided a dataset that contains the probability of working from home given the industrial sector (NAICS) an individual works in. The probabilities are reported in Tab.~\ref{tab:telework}.

\setlength{\tabcolsep}{9pt} 
\renewcommand{\arraystretch}{1} 
\begin{table}[!ht]
    \centering
    \begin{tabular}{llr}
\toprule
   CNS &                                        NAICS &  Teleworkability \\
\midrule
 CNS01 &         Agriculture, Forestry, Fishing and Hunting &          0.076394 \\
 CNS02 &      Mining, Quarrying, and Oil and Gas Extraction &          0.254480 \\
 CNS03 &                                          Utilities &          0.370015 \\
 CNS04 &                                       Construction &          0.185599 \\
 CNS05 &                                      Manufacturing &          0.224803 \\
 CNS06 &                                    Wholesale Trade &          0.517553 \\
 CNS07 &                                       Retail Trade &          0.143435 \\
 CNS08 &                     Transportation and Warehousing &          0.186145 \\
 CNS09 &                                        Information &          0.717062 \\
 CNS10 &                              Finance and Insurance &          0.762030 \\
 CNS11 &                 Real Estate and Rental and Leasing &          0.418109 \\
 CNS12 &   Professional, Scientific, and Technical Services &          0.802757 \\
 CNS13 &            Management of Companies and Enterprises &          0.791891 \\
 CNS14 &  Administrative and Support and Waste Managemen... &          0.310632 \\
 CNS15 &                               Educational Services &          0.826465 \\
 CNS16 &                  Health Care and Social Assistance &          0.252522 \\
 CNS17 &                Arts, Entertainment, and Recreation &          0.297494 \\
 CNS18 &                    Accommodation and Food Services &          0.035366 \\
 CNS19 &      Other Services (except Public Administration) &          0.312351 \\
 CNS20 &  Federal, State, and Local Government, excludin... &          0.414776 \\
\bottomrule
\end{tabular}
    \caption{Share of jobs that can be performed at home divided by their respective NAICS sector.}
    \label{tab:telework}
\end{table}

\newpage

\section{Sample composition}
Samples of users gathered as passively collected GPS trajectories from individuals' personal devices, such as mobile phones, are often afflicted by different biases making them not representative of a population. Nevertheless, it is possible, leveraging local demographic information, to partially remove and account for these biases getting closer to what could be a fair representation of a country's population. 

\subsection{Post-stratification reweighting: controlling biases in mobile phone data} \label{SIsec:reweighting}
As described in the manuscript, residential and workplace detection is of paramount importance in the inference of workers' industrial sectors. However, there is more that can be done to better exploit the joint information of GPS trajectories and demographic information available from census data. In particular, when analyzing GPS location data, it has become more and more clear the importance of de-biasing personal device data from uneven user base distributions~\cite{yabe2023behavioral, lucchini2023socioeconomic}. 
To this end, we leverage demographic information at the census block group level to reconstruct a population-representative base of individuals. This is performed by an individual-based reweighting step which leverages the detection of the individual residential areas (e.g., see \cite{yabe2023behavioral, lucchini2023socioeconomic}). Representativity is ensured by using state-specific stratified population information and by reweighting upsampled individuals from census block groups where at least one individual in our data is residing based on the specific strata population. Single individual weights are assigned based on the fraction of the stratified population of each census block group.

The procedure consists of the following steps: i) single individuals are upsampled to enrich our dataset; ii) upsampled individuals are assigned, based on demographic and industry sector, information about their probable sex, age group, income level, race, education level, and industrial sector; iii) individual weights are independently assigned for each of the demographic and work sector information based on the strata distribution at the state level; iv) weights are used for bootstrapping to construct a synthetic representative population consisting of $500,000$ individuals for each state included in the study. For population-wide analyses, each individual is treated in our analysis as an independent data point. In this situation, stop location sequences are aggregated together, providing a representative picture of employed/unemployed behavioural differences. In contrast, strata-specific results report group-specific behaviour whose representativity is controlled for by computing errors exclusively including originally independent stop-location sequences.

Additionally, to extend the validity of our results we aggregate results from multiple states displaced across different geographical areas of the US selecting them based on the internal repartition of their workforce into the three main economic sectors (see SI \ref{SIsec:state_selection}).

\subsection{States selection}\label{SIsec:state_selection}
\begin{figure}[ht]
\centering
\includegraphics[width=\linewidth]{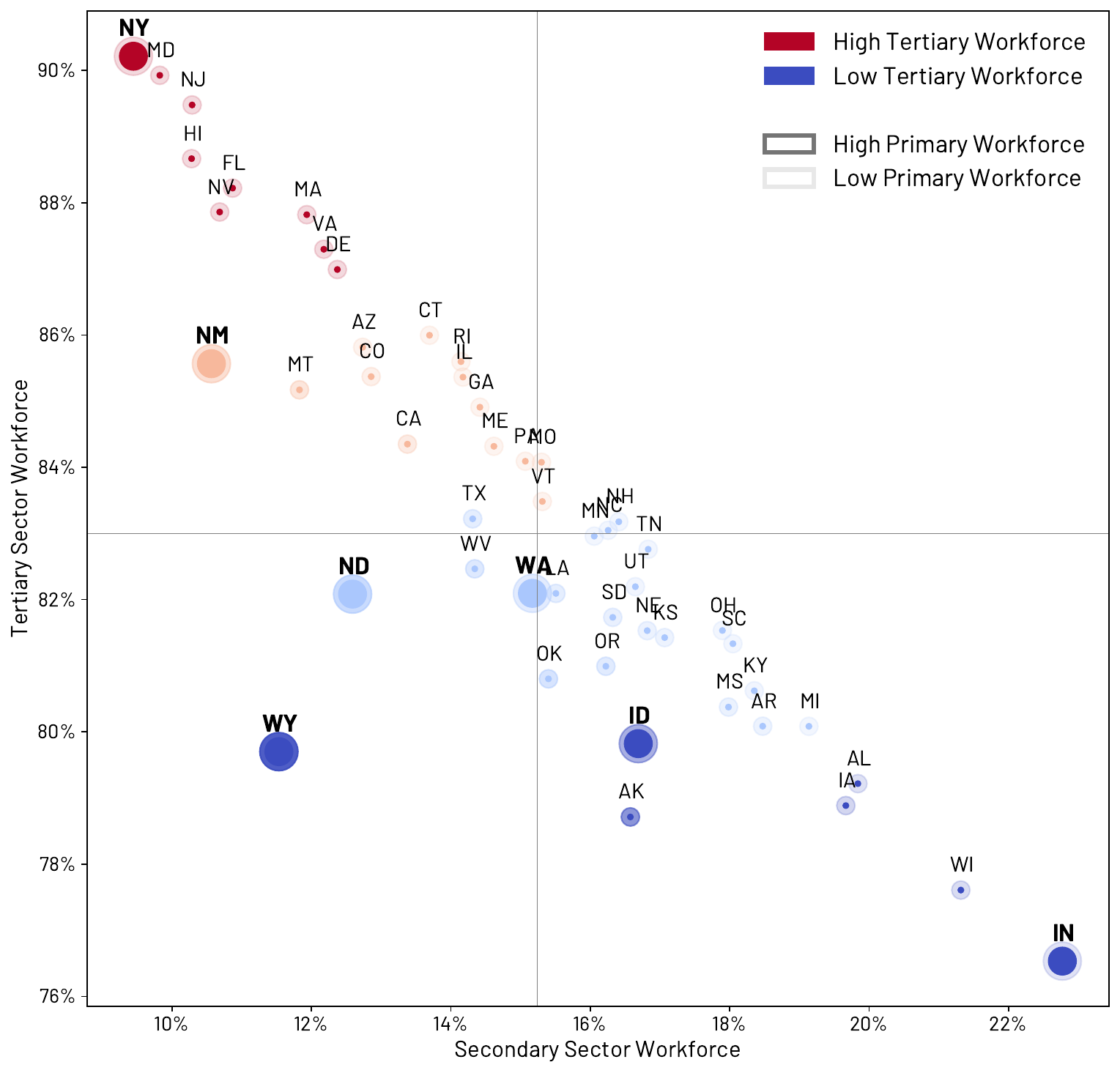}
\caption{State selection based on the primary, secondary, and tertiary workforce composition in the US states. The selected states are New York (NY), Wyoming (WY), Indiana (IN), Idaho (ID), Washington (WA), North Dakota (ND), and New Mexico (NM). The internal colour code distinguishes states with a high-tertiary workforce (in red) from those with a low-tertiary workforce (in blue). The intensity of the shaded circle surrounding states highlights those states with higher levels of the primary workforce (higher intensity corresponds to higher primary workforce levels).}
\label{fig:states_selection}
\end{figure}

\begin{figure}[ht]
\centering
\includegraphics[width=0.9\linewidth]{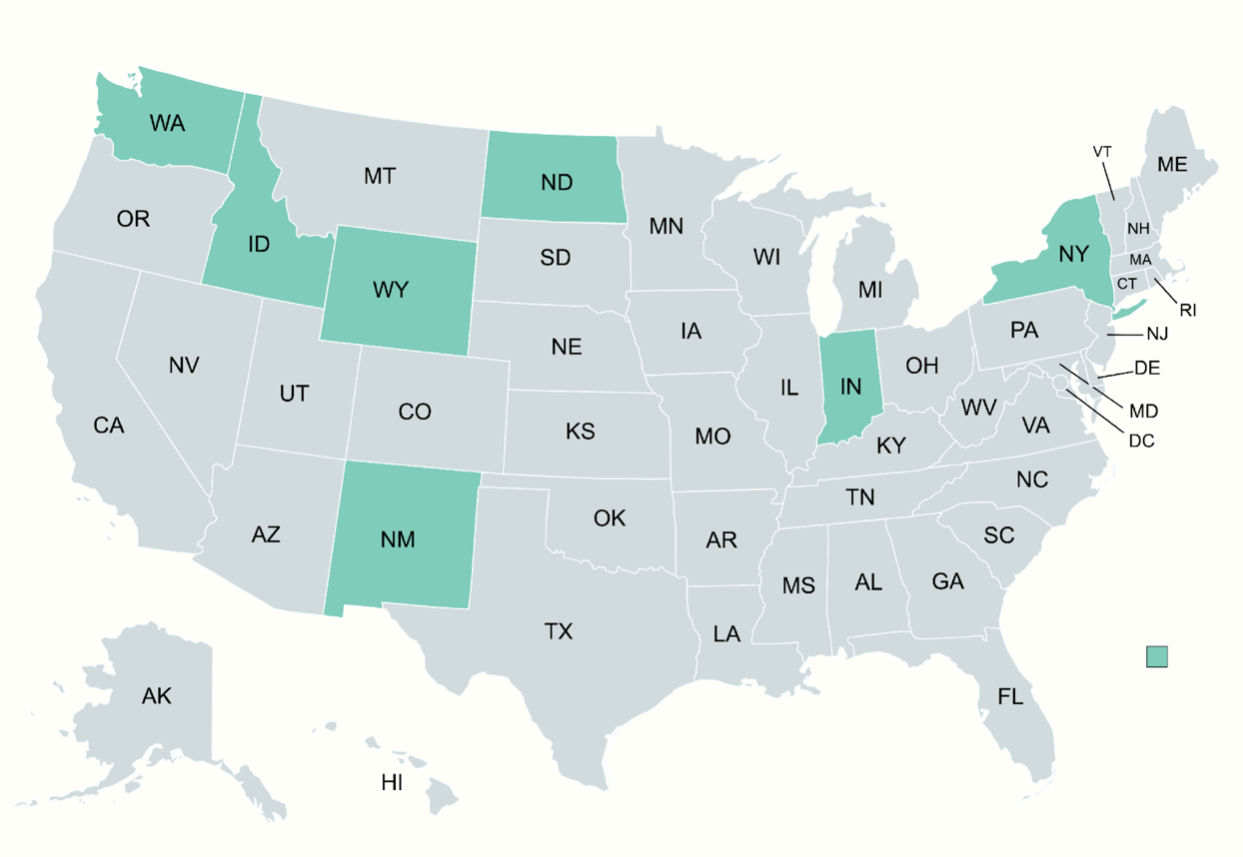}
\caption{Geographical displacement of the countries included in this work analysis.}
\label{fig:states_geo_distribution}
\end{figure}

To further improve the data representation we focus our analysis on seven US states: New York, Wyoming, Indiana, Idaho, Washington, North Dakota, and New Mexico. These states have been selected to take into account at the same time:
\begin{itemize}
    \item the representation of different workforce compositions in terms of population distribution across primary, secondary and tertiary economic sectors. Fig.~\ref{fig:states_selection} shows the distribution of the states taking into account the percentages of the primary, secondary and tertiary workforce in each US states. We selected the states in order to cover the entire space. The information about the economics sector are derived from the NAICS sector data and mapped according to Tab.~\ref{tab:naics_economic_sector}; 
    \item the geographical representation of the states to avoid the selection of states the belongs to a single geographic region (e.g., Northeast, etc.). Fig.~\ref{fig:states_geo_distribution} depicts the selected states' geographical distribution.
\end{itemize}

\begin{table}[!hbt]
\resizebox{\textwidth}{!}{%
\begin{tabular}{@{}lll@{}}
\toprule
\textbf{NAICS sector} & \textbf{Description}                                                     & \textbf{Economic sector} \\ \midrule
11    & Agriculture, Forestry, Fishing and Hunting    & Primary   \\
21    & Mining, Quarrying, and Oil and Gas Extraction & Primary   \\
22    & Utilities                                     & Secondary \\
23    & Construction                                  & Secondary \\
31-33 & Manufacturing                                 & Secondary \\
42    & Wholesale Trade                               & Tertiary  \\
44-45 & Retail Trade                                  & Tertiary  \\
48-49 & Transportation and Warehousing                & Tertiary  \\
51    & Information                                   & Tertiary  \\
52    & Finance and Insurance                         & Tertiary  \\
53    & Real Estate and Rental and Leasing            & Tertiary  \\
54                    & Professional, Scientific, and Technical Services                         & Tertiary                 \\
55    & Management of Companies and Enterprises       & Tertiary  \\
56                    & Administrative and Support and Waste Management and Remediation Services & Tertiary                 \\
61    & Educational Services                          & Tertiary  \\
62    & Health Care and Social Assistance             & Tertiary  \\
71    & Arts, Entertainment, and Recreation           & Tertiary  \\
72    & Accommodation and Food Services               & Tertiary  \\
81    & Other Services (except Public Administration) & Tertiary  \\
92    & Public Administration                         & Tertiary  \\ \bottomrule
\end{tabular}%
}
\caption{NAICS sectors and their corresponding economic sectors.}
\label{tab:naics_economic_sector}
\end{table}

\clearpage

\section{Job loss detection}
\subsection{Remote work adjustment}\label{SIsec:remote_adjust}
    In this section, we explain in greater detail the procedure to assign to each sector at each time $t$ a probability of working remotely (or being unemployed). Following the idea described in the main text, we aim to properly define a probability that accounts for both the amount of work that can still be performed from remote, $\tilde{R}_s(t)$, and the reduction work that was performed in person during the baseline period (which is either now performed by employees working remotely or not performed anymore, thus making those individual unemployed), $\tilde{r}_s(t)$. Both these quantities express sector-specific aggregate measures to avoid further assumptions, needed, for example, in case we would want to build an individual probability measure.

    The probability presented in the main text ($P_s(t)$, we report the definition below) uses the ratio between the remaining work time that could be performed from remote and the reduction in in-presence work time as a conservative measure, likely to undershoot the probability of being unemployed. 
    \begin{equation*}
        P_s(t) = 1-\min\Big(\frac{\tilde{R}_s(t)}{\tilde{r}_{s}(t)}, 1\Big)
    \end{equation*}
    The measure is bounded between $0$, in the case where $\tilde{r}_s(t) \le \tilde{R}_s(t)$ (i.e. when the amount of time that can still be worked from remote exceeds the reduction of in-person work time), and $1$, in the case where $\tilde R_s(t) = 0$ (i.e. when all the remote-work time is already accounted for by ``not at risk'' individuals, those with $rt_s(t)<1$).
    The aim of this section is to introduce the formal definition of both quantities ($\tilde R_s(t)$ and $\tilde r_s(t)$) and to provide a more clear understanding of how those quantities are computed. To this end, we report here the definitions introduced in the main text of:
    \begin{equation*}
    \begin{cases}
        r_u(t) = 1 & \text{if }v_u(t) = 0\\
        r_u(t) = 0 & \text{otherwise}
    \end{cases}
    \end{equation*}
    (being $v_u(t)$ the number of visits $u$ made to their work location within the time window $t$), and
    \begin{equation*}
        rt_u(t) = 1 -\frac{w_u(t)}{w_u(b)}.
    \end{equation*}
    with $w_u(t)$ being the time the user $u$ spent at the workplace in the time window $t$, and $w_u(b)$ being the median of the time spent at work (within windows of the same size as $t$) during the baseline period (Jan 3 - Mar 7, 2020).
    
    \subsubsection{Adjusted sector remote work fraction: $\tilde R_s(t)$}
    
    $\tilde R_s(t)$ is defined for each time window $t$ by the following formula:
    $$\tilde R_s(t) = \max\Bigg(0, R_s-W_s(t)\Bigg);$$
    where $R_s$ is the fraction of work that can be performed remotely, and $W_s(t)$ is the weighted fraction of work performed at home by individuals who are still commuting at time $t$. Formally, $R_s$ is a quantity taken from the literature for each different NAICS (see Dingel et al.~\cite{dingel2020many}), while $W_s(t) = \sum\limits_{u' \in s} rt_u'(t) * rw_{u_i}'$. Here $rw_u$ is an individual-specific weight. It is proportional to the fraction of time the individual was spending at work with respect to the amount of time spent at work by the entire sector workforce during the baseline period. Formally it is defined as: 
    $$rw_u = \frac{\bar{w}_u(b)}{\sum\limits_{u' \in s}\bar{w}_{u'}(b)}$$
    where $\bar w_u(b)$ is the average time spent at work in a day by a single individual during the baseline period $b$.

    \subsubsection{Weighted reduction in in-person work time: $\tilde r_s(t)$}
    Combining $r_u(t)$ and $rt_u(t)$ with $rw_u$, as defined above we can formalize the concept of ``weighted fraction of remaining work time that could be performed remotely by those individuals who stopped visiting their work location''. The decision whether this time will correspond effectively to remote work time is encoded in the formula for the unemployment probability, here we focus on formalizing the $\tilde r_s(t)$ mathematical definition:
    $$\tilde{r}_{s}(t)=\sum\limits_{u' \in s} rt_{u'}(t) * rw_{u'} - \sum\limits_{u' \in s} (1 - r_{u'}(t)) * rt_{u'}(t) * rw_{u_i}.$$
    The first term captures the sector-specific weighted reduction in the amount of time spent at work by the entire workforce population within each time window $t$. The second term represents the weighted reduction of time spent at work by those individuals who are still going to their workplace in $t$. Thus, the difference between the two represents the total weighted reduction in time spent at work in $t$ for those individuals. 
    A complementary definition would be to directly refer to the weighted reduction of ``at-risk'' individuals: $\tilde{r}_{s}(t)=\sum\limits_{u' \in s} r_{u'}(t) * rw_{u_i}.$
    The efficacy of this procedure is to easily account for systemic reductions in the in-presence work time alongside the intrinsic industrial sector remote workability.
    
    \subsubsection{Remote workability mechanism: an intuition}

    The idea behind this procedure is to use $rt_u(t) * rw_u$ to re-weight the risk of unemployment for each user $u$ in each time window $t$ to take into account the fraction of visits/work time that particular user was ``consuming'' during the baseline compared to all other users in a specific window and within the same NAICS. 
    Intuitively, by aggregating $rt_u(t) * rw_u$ at the sector level for all users that are still visiting their work location ($r_u(t) < 1$), we are computing a weighted average (over the fraction of time each user worked) of the reduction of the time spent at work ($rt_u(t)$). This can also be seen as the change in remote work that we measure from a reduction in the time spent at work by users that are still visiting their work location, i.e. a measure of how much remote work is already performed by individuals who are still employed.

    \begin{figure}[!htb]
    \centering
    \includegraphics[width=\linewidth]{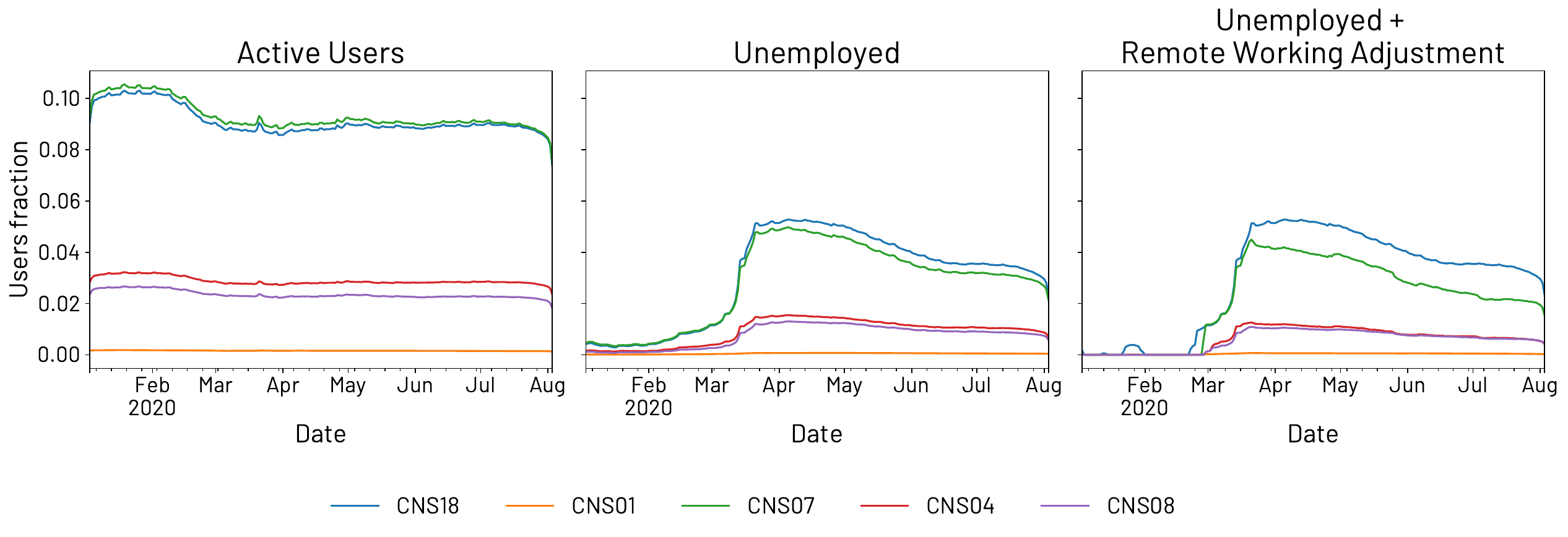}
    \caption{Remote working mechanism with NAICS sectors that are less teleworkable: Accommodation and Food Services	(CNS18); Agriculture, Forestry, Fishing and Hunting (CNS01); Retail Trade (CNS07); Construction (CNS04); and Transportation and Warehousing (CNS08). Given the less remote workability, the adjustment leaves the unemployed unaffected (middle and left panel). The share of jobs that can be performed at home are reported in Tab. \ref{tab:telework}.}
    \label{fig:less_teleworkable}
    \end{figure}
    
    \begin{figure}[!htb]
    \centering
    \includegraphics[width=\linewidth]{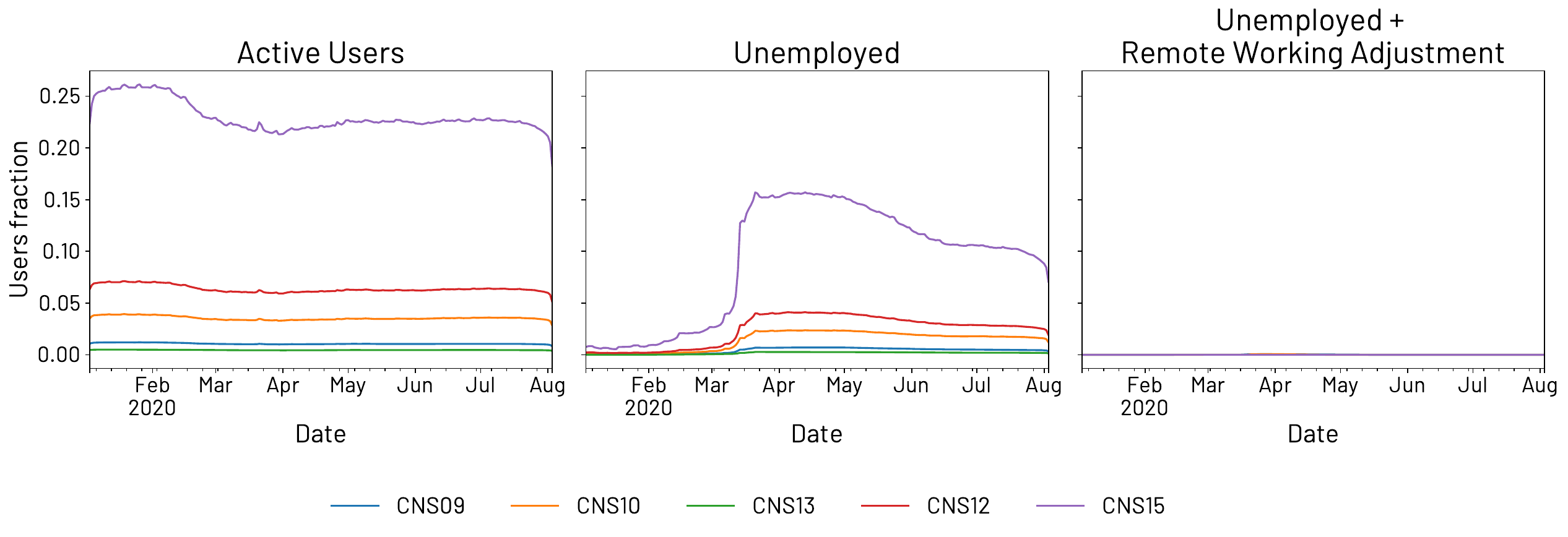}
    \caption{Remote working mechanism with NAICS sectors that are more teleworkable: Educational Services (CNS15); Professional, Scientific, and Technical Services (CNS12); Management of Companies and Enterprises (CNS13); Finance and Insurance (CNS10); and Information (CNS09). Given the more teleworkability the adjustment  ``removes" the unemployed since they are working from home (middle and left panel). The share of jobs that can be performed at home are reported in Tab. \ref{tab:telework}.}
    \label{fig:more_teleworkable}
    \end{figure}

\newpage

\section{Algorithm evaluation}
In this section, we detail the performance of our unemployment detection algorithm. Given the differences in data sources between Unemployment Insurance (UI) claim counts and the official unemployment figures from the Local Area Unemployment Statistics (LAUS) program (see Section~\ref{SI:sec:datasets}), we evaluate the algorithm using both datasets. We chose these datasets for their differing temporal and spatial resolutions:
\begin{itemize}
    \item \textit{Temporal Resolution}: UI claims data are reported weekly, providing a more immediate, near real-time basis for evaluating algorithm performance. By contrast, LAUS data represent official unemployment figures calculated through an estimation process, incorporating multiple sources (including UI claims) and are published with a lag.
    \item \textit{Geographic Resolution}: UI claims are available at the state level, segmented by NAICS sectors, which allows sector-specific performance analysis. Meanwhile, LAUS data support a finer geographic resolution, enabling us to evaluate our algorithm at the county level.
\end{itemize}

\begin{figure}[!htb]
\centering
\includegraphics[width=0.8\linewidth]{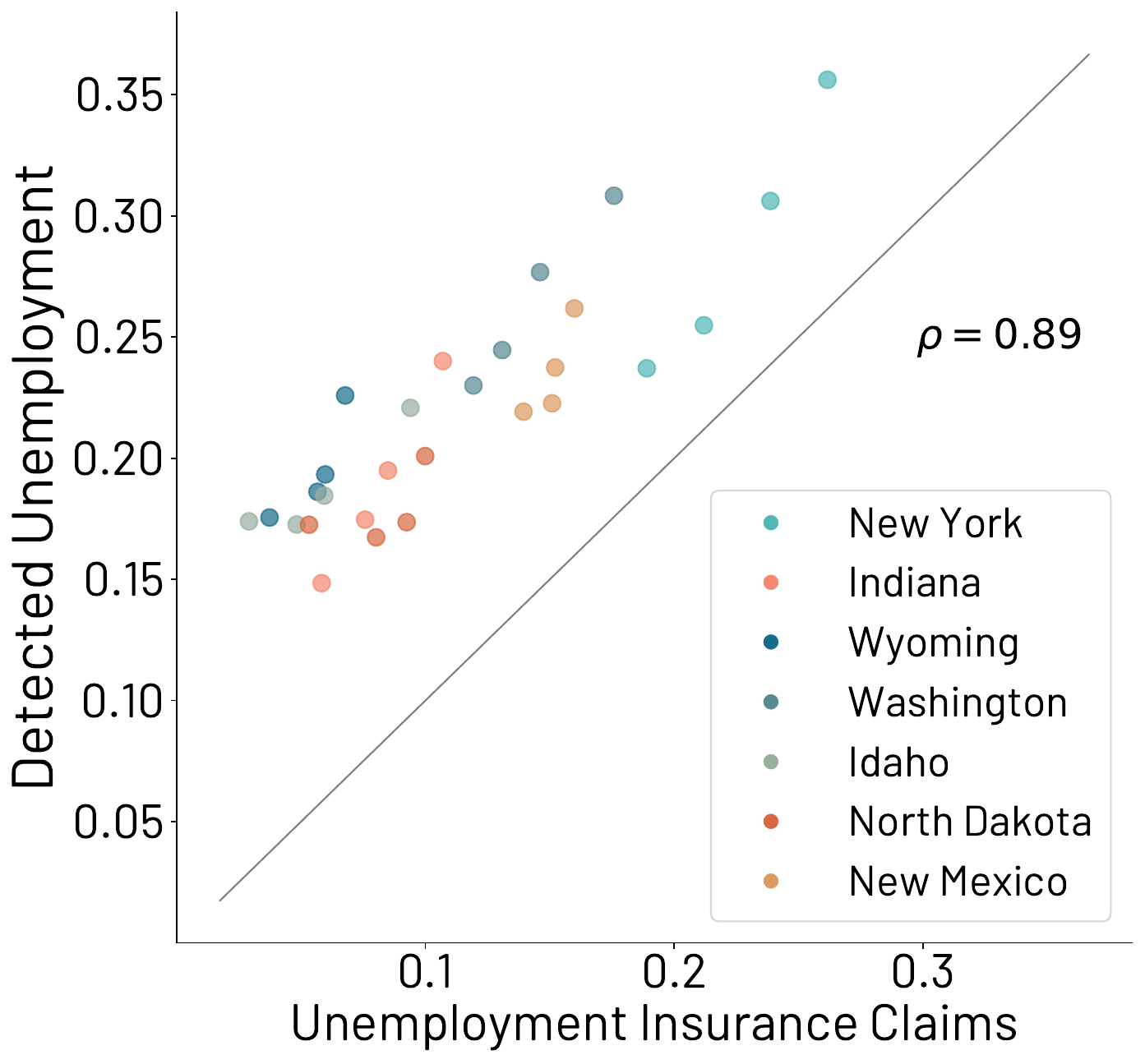}
\caption{Algorithm evaluation and average Pearson correlation using the Unemployment Insurance (UI) claims data for each state and month.}
\label{fig:eval_state_months_uiclaims}
\end{figure}

\begin{figure}[!htb]
\centering
\includegraphics[width=\linewidth]{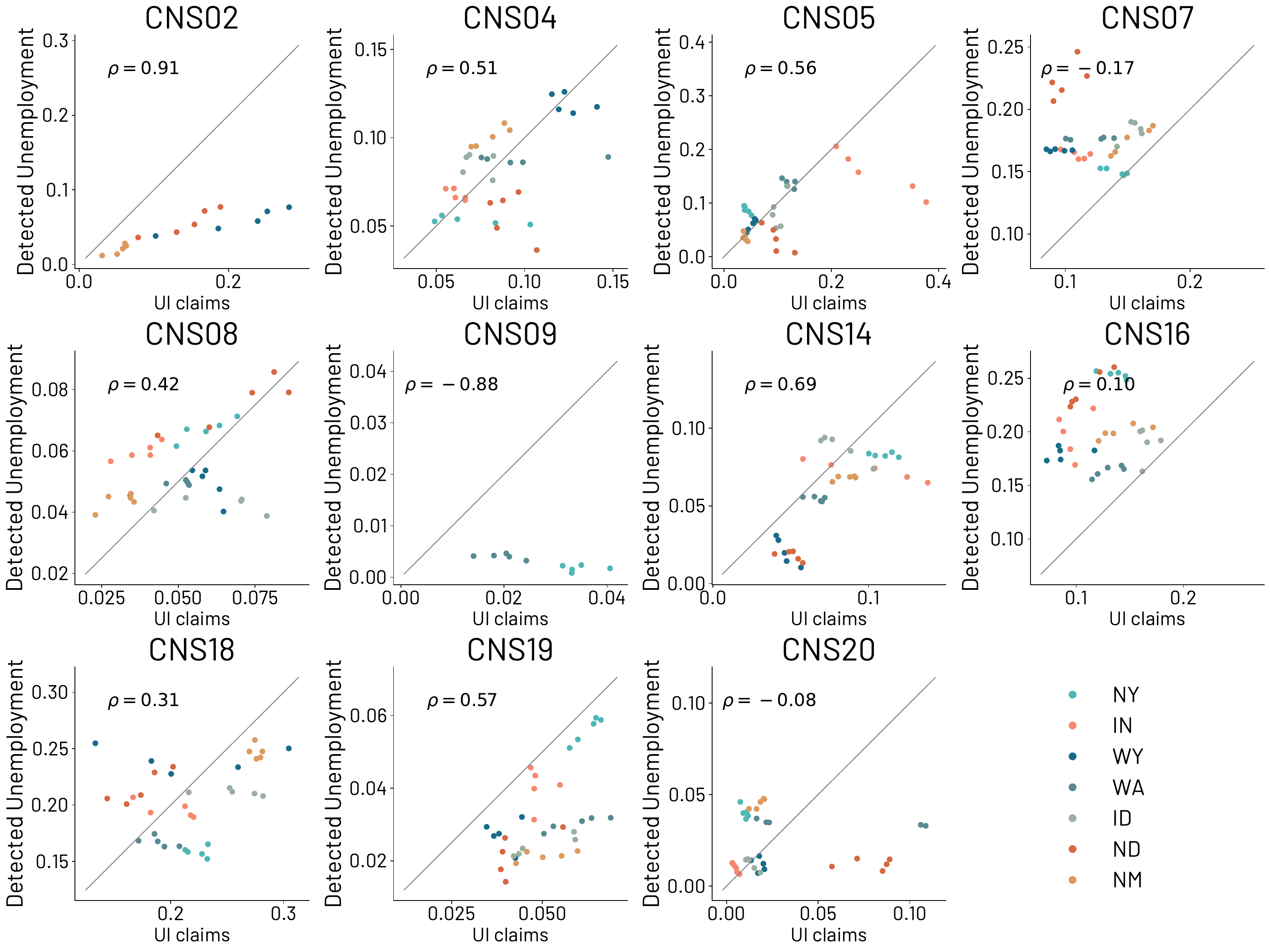}
\caption{Algorithm evaluation and average Pearson correlation using the Unemployment Insurance (UI) claims data in each NAICS sector for each state and month. Only representative NACIS are showed.}
\label{fig:eval_cns_months_uiclaims}
\end{figure}

\begin{figure}[!htb]
\centering
\includegraphics[width=0.7\linewidth]{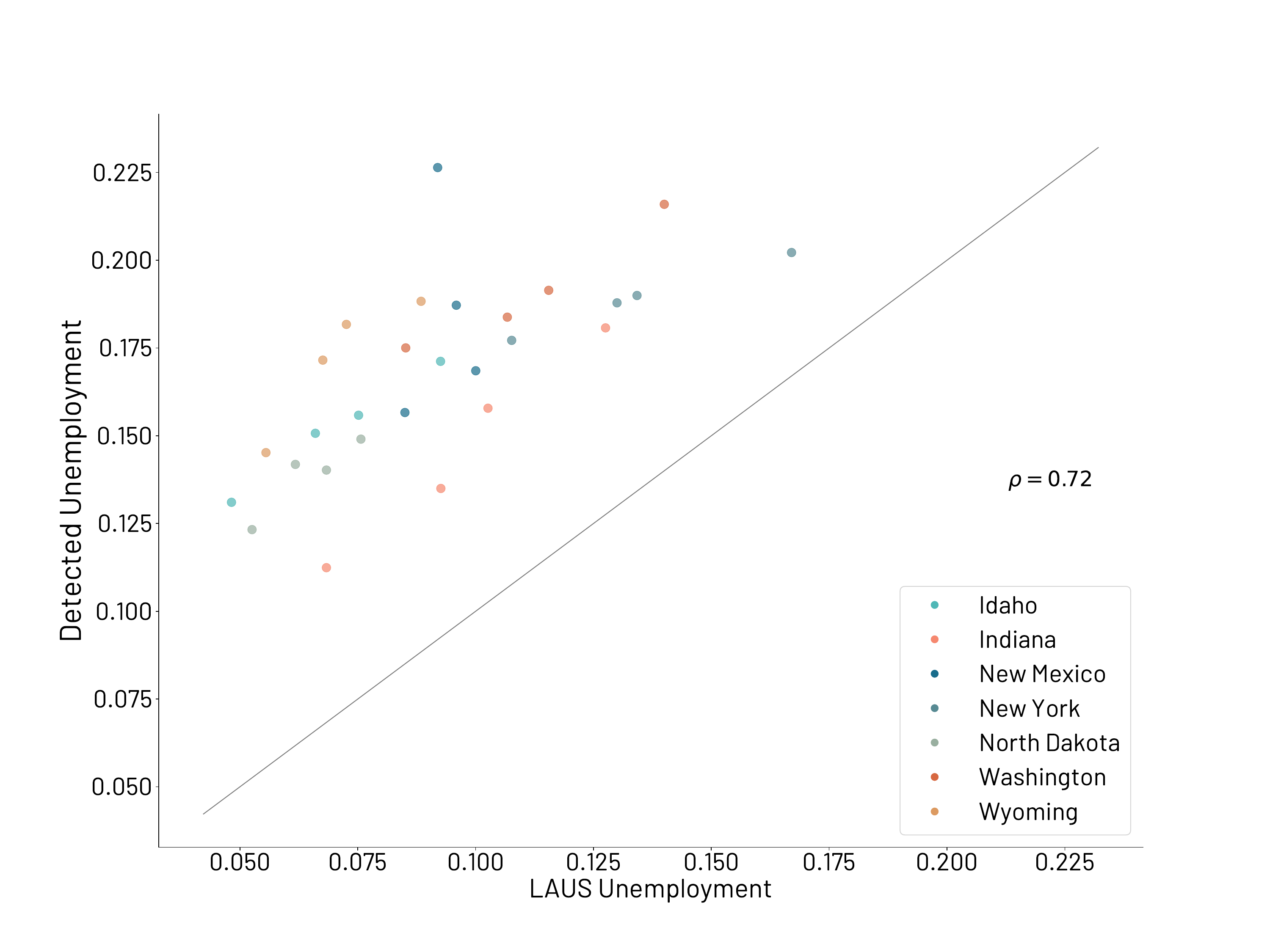}
\caption{Algorithm evaluation and average Pearson correlation using the Local Area Unemployment Statistics (LAUS) data across all states and month.}
\label{fig:laus_mean_eval}
\end{figure}

\begin{figure}[!htb]
\centering
\includegraphics[width=0.7\linewidth]{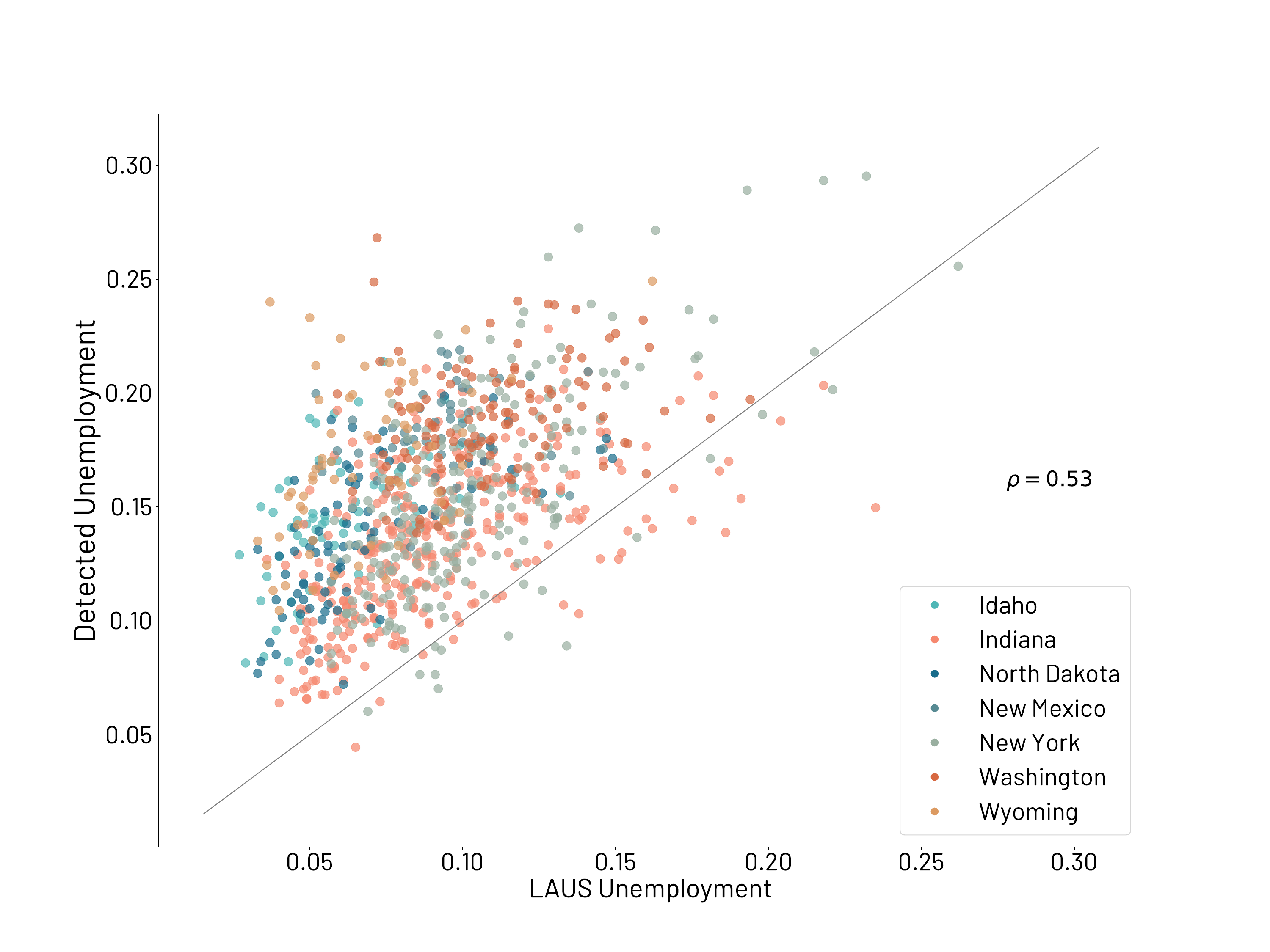}
\caption{Algorithm performance evaluation and average Pearson correlation using Local Area Unemployment Statistics (LAUS) data across all counties and months.}
\label{fig:laus_county_eval}
\end{figure}

\begin{figure}[!htb]
\centering
\includegraphics[width=\linewidth]{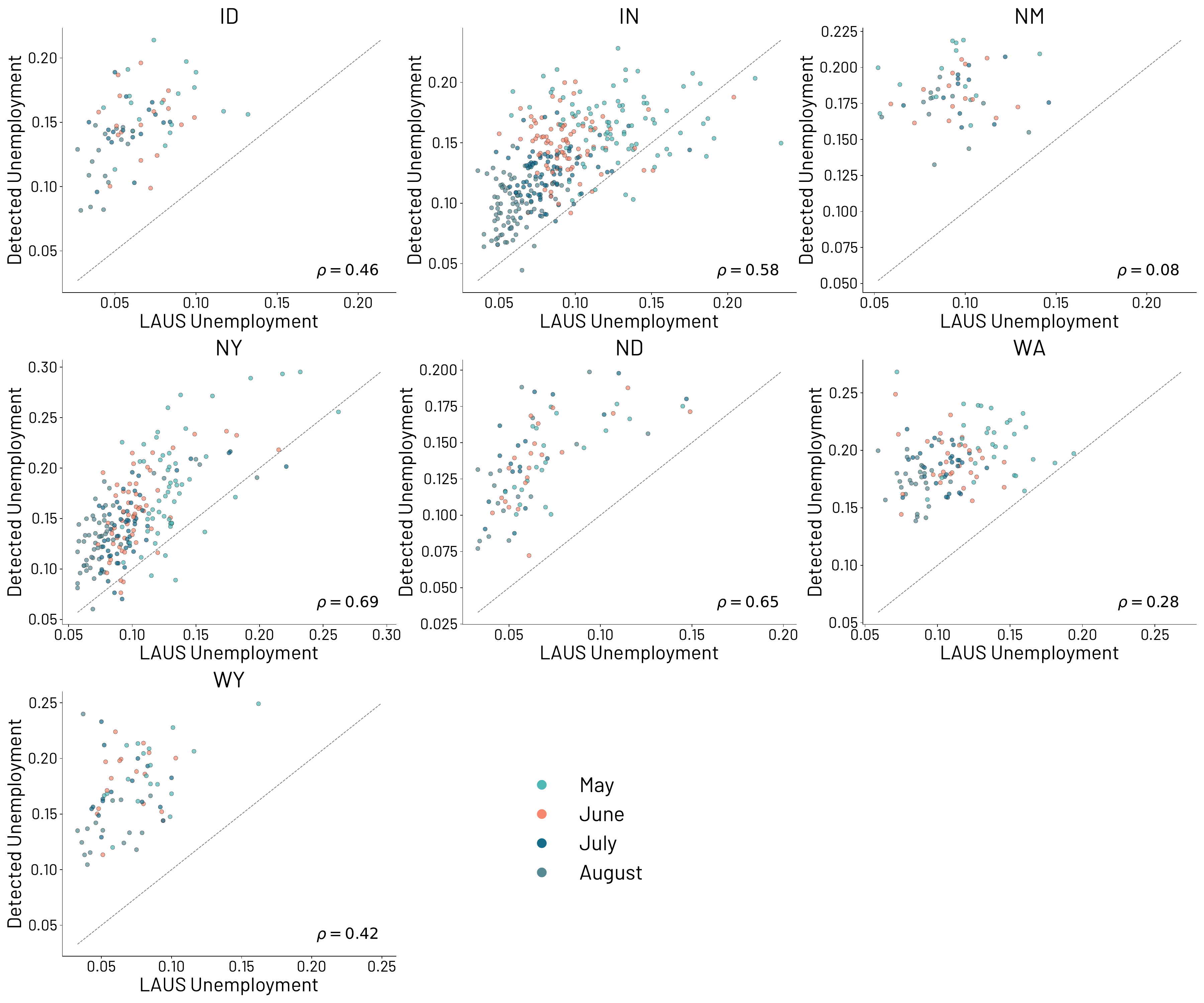}
\caption{Algorithm performance evaluation and average Pearson correlation using Local Area Unemployment Statistics (LAUS) data across all counties divided by month.}
\label{fig:laus_county_month_eval}
\end{figure}

To assess accuracy, we calculate the Pearson correlation coefficient between the monthly unemployment rates estimated by our algorithm and the corresponding unemployment rates from the UI claims at the state level ($\rho = 0.89$, see Fig.~\ref{fig:eval_state_months_uiclaims}). We further provide sector-specific correlations for NAICS sectors (see Fig.~\ref{fig:eval_cns_months_uiclaims}). Using the LAUS data, we calculate the Pearson correlation between our algorithm’s monthly unemployment rate estimates and the official LAUS unemployment rates at the state level ($\rho = 0.72$, see Fig.~\ref{fig:laus_mean_eval}). Additionally, we present finer-grained correlations at the county level (see Fig.~\ref{fig:laus_county_eval}) and provide monthly-level correlations (see Fig.~\ref{fig:laus_county_month_eval}).

\section{Demographic differences}
In this section we disaggregate the mobility behaviour of unemployed individuals ($t > 0$) based on the individual’s socio-demographic group. Socio-demographic information is extracted from the Longitudinal Employer-Household Dynamics (LODES) dataset and includes Sex, Age, Income, Race, and Education. 
We provide, for each mobility metric, namely radius of gyration (SI ~\ref{sec:radius}), entropy (SI ~\ref{sec:entropy}) and capacity (SI ~\ref{sec:capacity}), the z-scores of the metric over time and their distributions divided by a specific socio-demographic indicator.
Moreover, for each mobility metric, we provide statistical tests for (i) the comparison of the groups of a socio-demographic against the entire employed population (SI Tab. \ref{tab:unemployed_vs_all_empl_population}); (ii) the comparison of the groups of a socio-demographic against their corresponding employed population (e.g., male unemployed vs male employed) (SI Tab. \ref{tab:unempl_empl_same_demographic_group}); and (iii) the comparison to test whether there are differences between groups inside the same socio-demographic factor (e.g., male unemployed vs female unemployed) (SI Tab. \ref{tab:unemployed_within_demographic_group}).

\subsection{After job loss demographic differences: Radius of gyration}\label{sec:radius}

\begin{figure}[!htb]
\centering
\includegraphics[width=\linewidth]{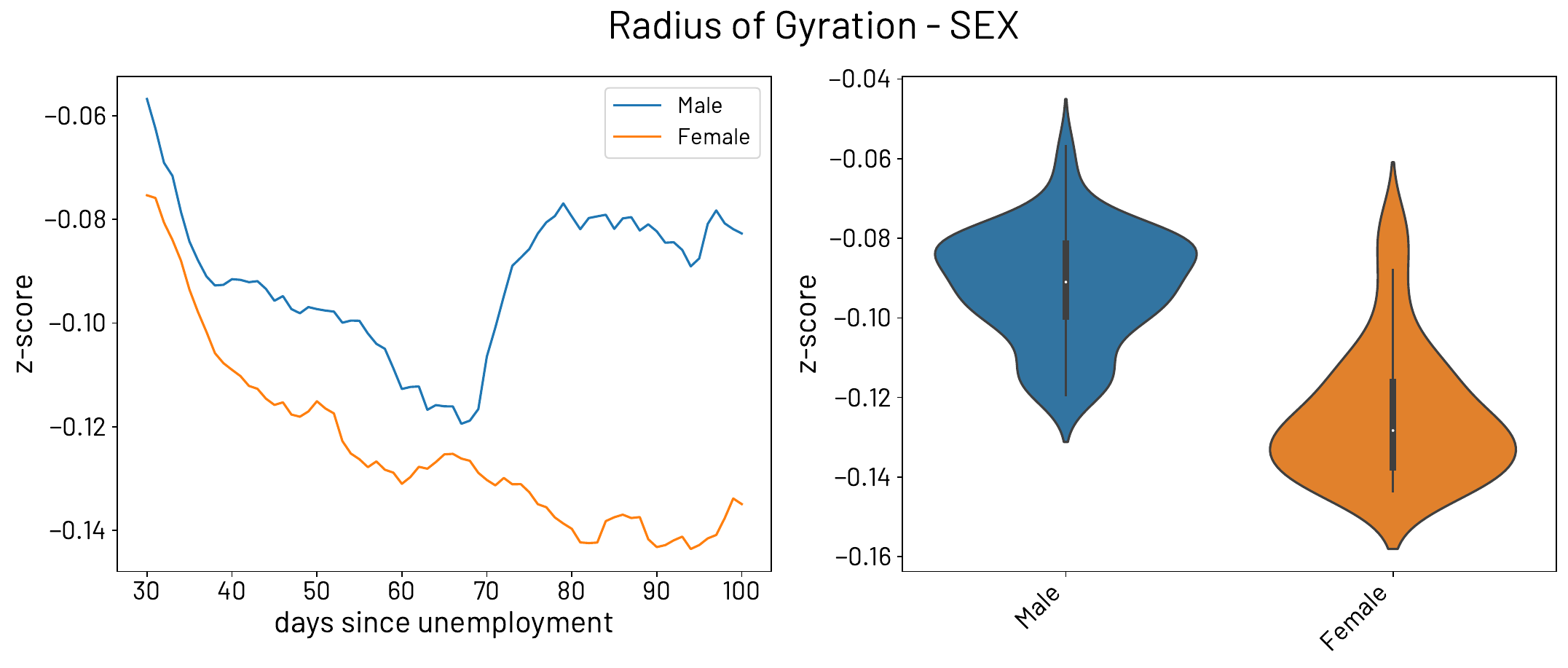}
\caption{Z-scores for the radius of gyration divided by the Sex demographic indicator and their corresponding distribution.}
\label{fig:decouple_sex_radius}
\end{figure}

\begin{figure}[!htb]
\centering
\includegraphics[width=\linewidth]{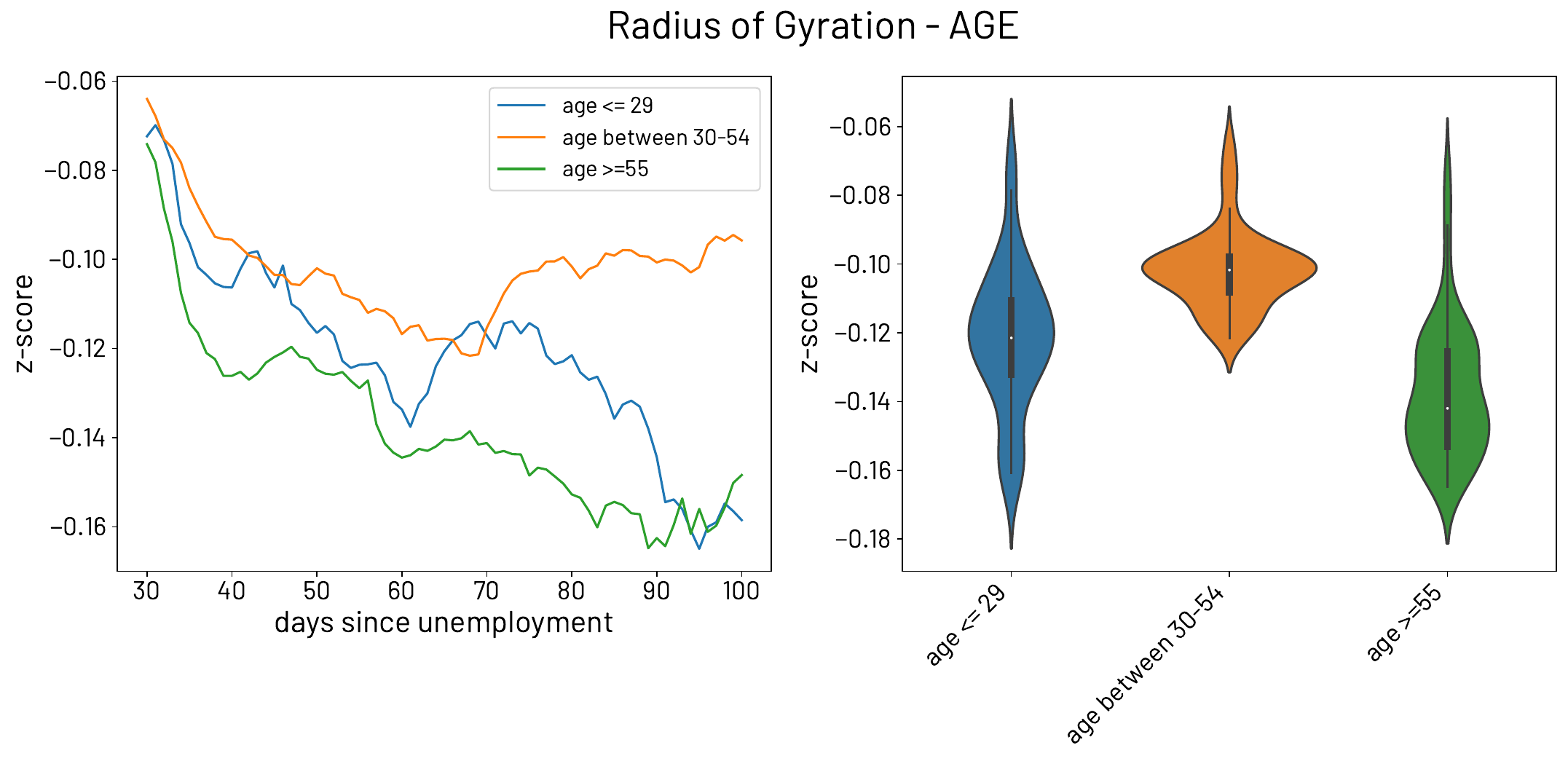}
\caption{Z-scores for the radius of gyration divided by the Age demographic indicator and their corresponding distribution.}
\label{fig:decouple_age_radius}
\end{figure}

\begin{figure}[!htb]
\centering
\includegraphics[width=\linewidth]{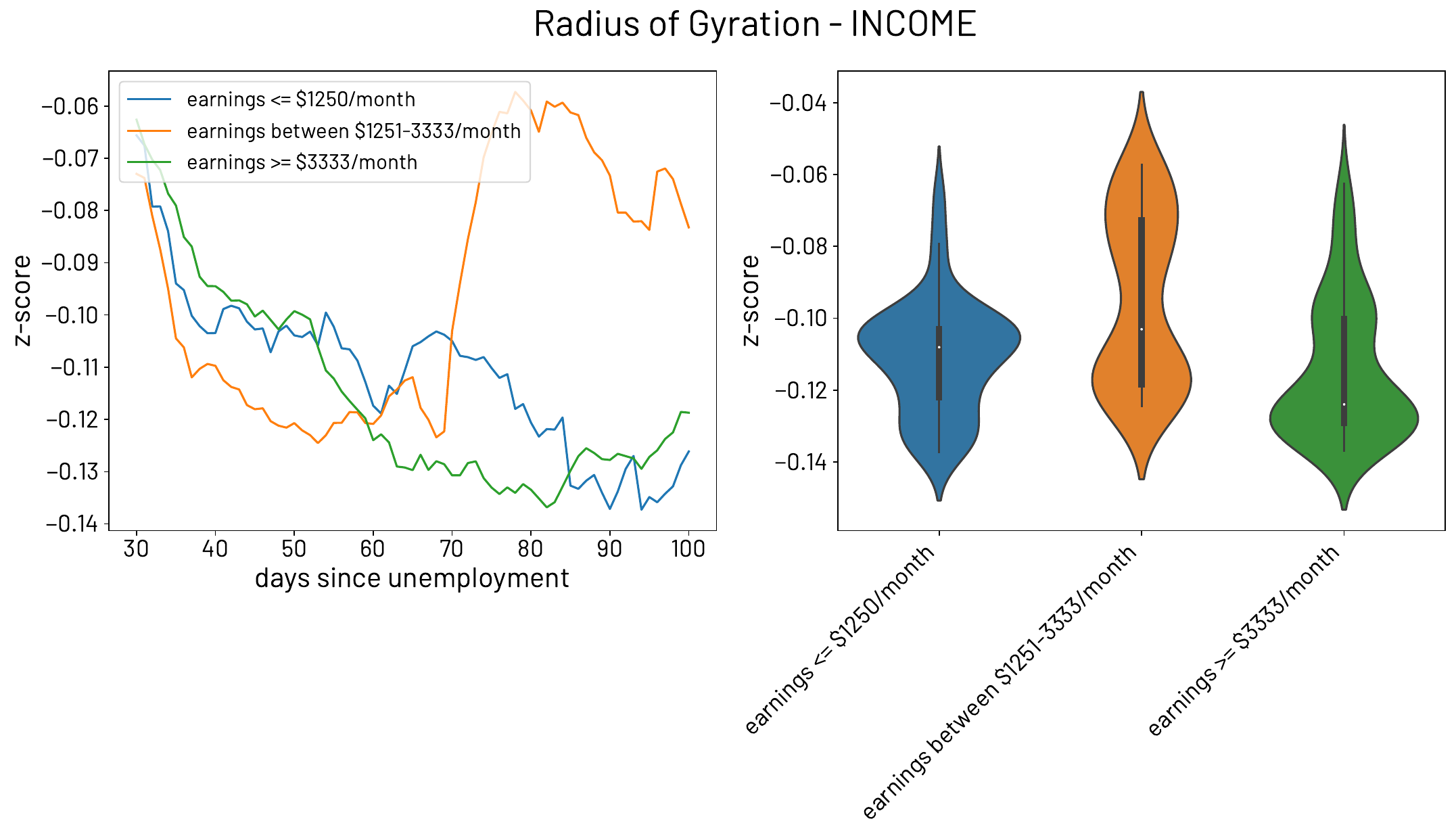}
\caption{Z-scores for the radius of gyration divided by the Income demographic indicator and their corresponding distribution.}
\label{fig:decouple_income_radius}
\end{figure}

\begin{figure}[!htb]
\centering
\includegraphics[width=\linewidth]{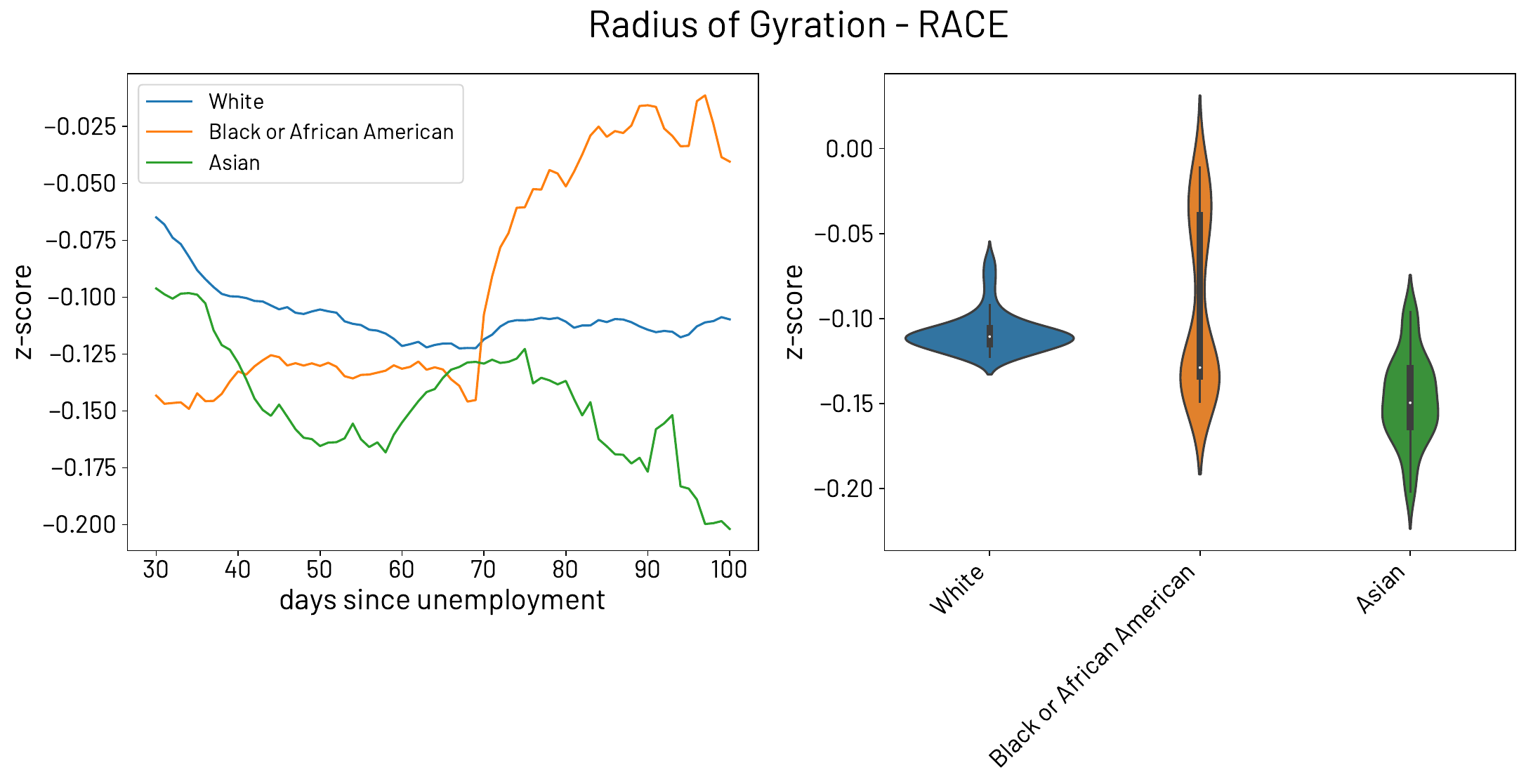}
\caption{Z-scores for the radius of gyration divided by the Race demographic indicator and their corresponding distribution.}
\label{fig:decouple_race_radius}
\end{figure}

\begin{figure}[!htb]
\centering
\includegraphics[width=\linewidth]{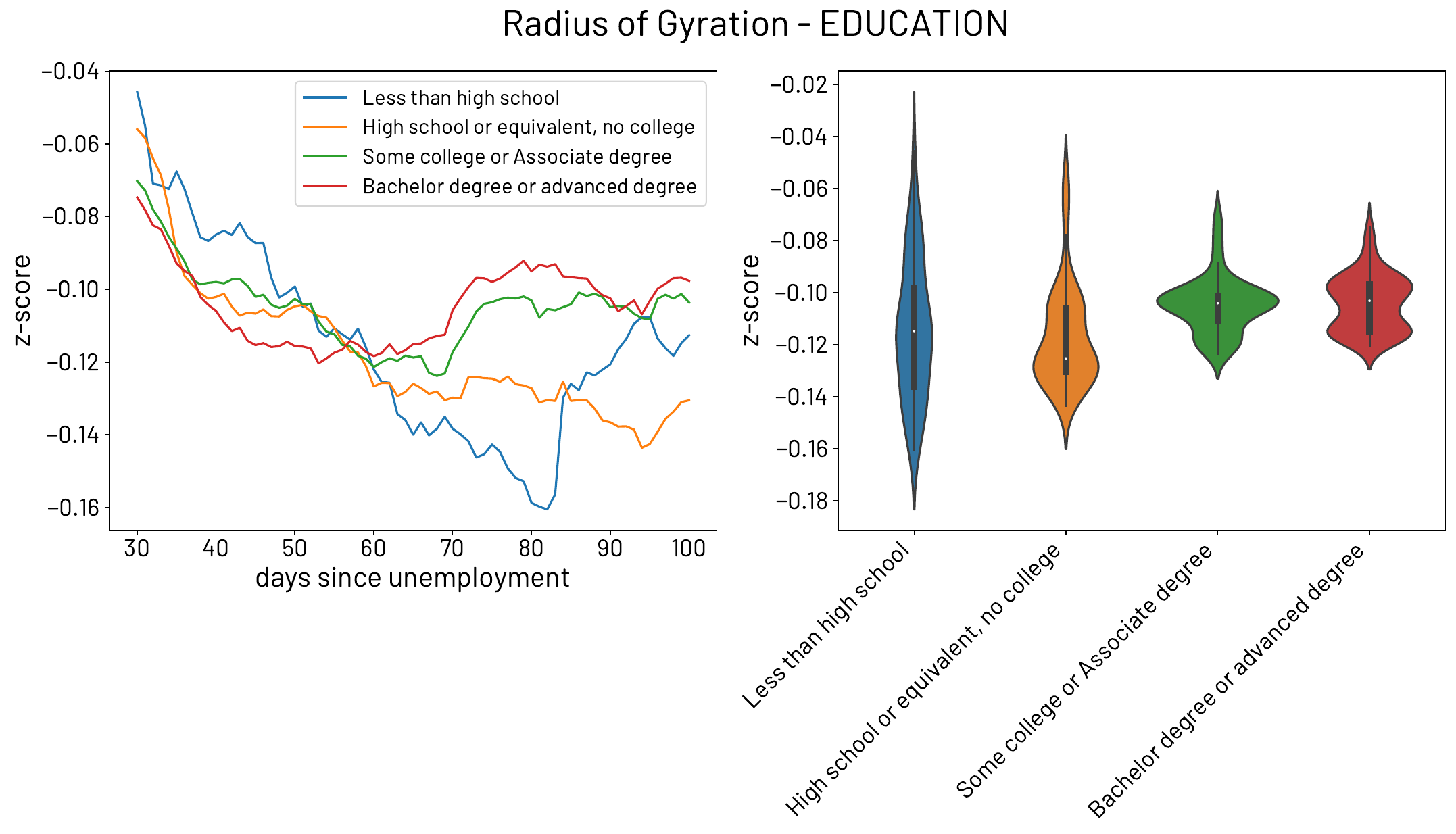}
\caption{Z-scores for the radius of gyration divided by the Education demographic indicator and their corresponding distribution.}
\label{fig:decouple_education_radius}
\end{figure}

\subsection{After job loss demographic differences: Entropy}\label{sec:entropy}

\begin{figure}[!htb]
\centering
\includegraphics[width=\linewidth]{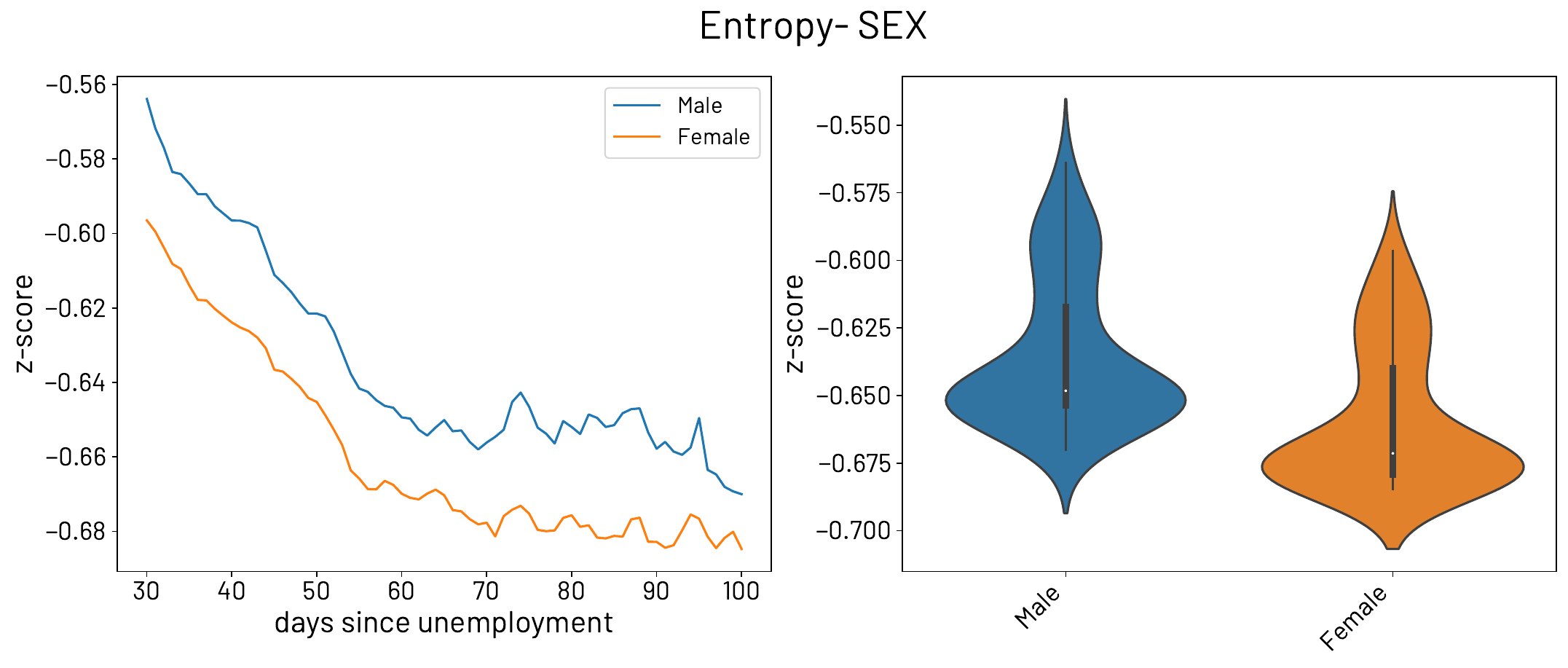}
\caption{Z-scores for the entropy divided by the Sex demographic indicator and their corresponding distribution.}
\label{fig:decouple_sex_entropy}
\end{figure}

\begin{figure}[!htb]
\centering
\includegraphics[width=\linewidth]{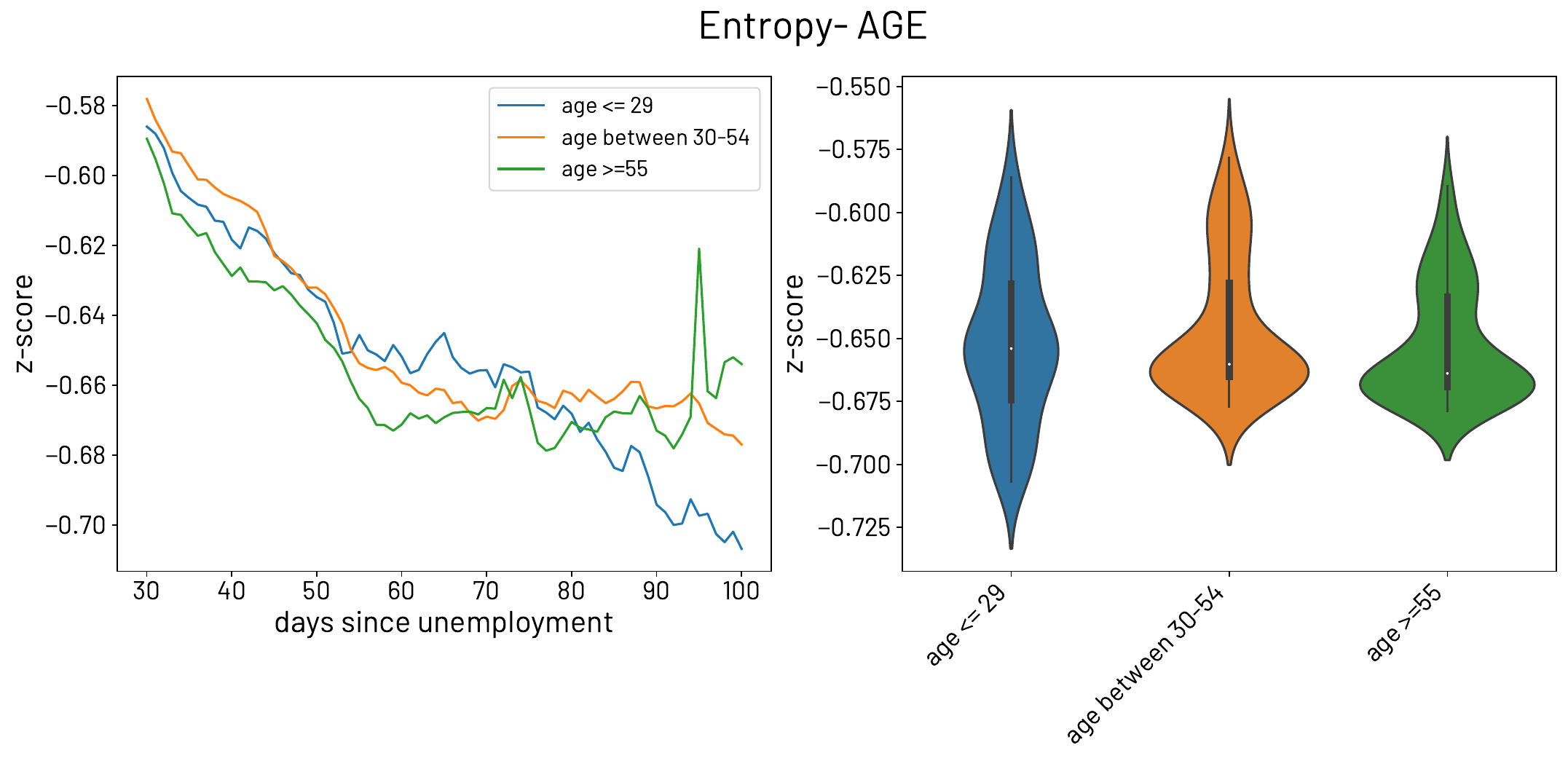}
\caption{Z-scores for the entropy divided by the Age demographic indicator and their corresponding distribution.}
\label{fig:decouple_age_entropy}
\end{figure}

\begin{figure}[!htb]
\centering
\includegraphics[width=\linewidth]{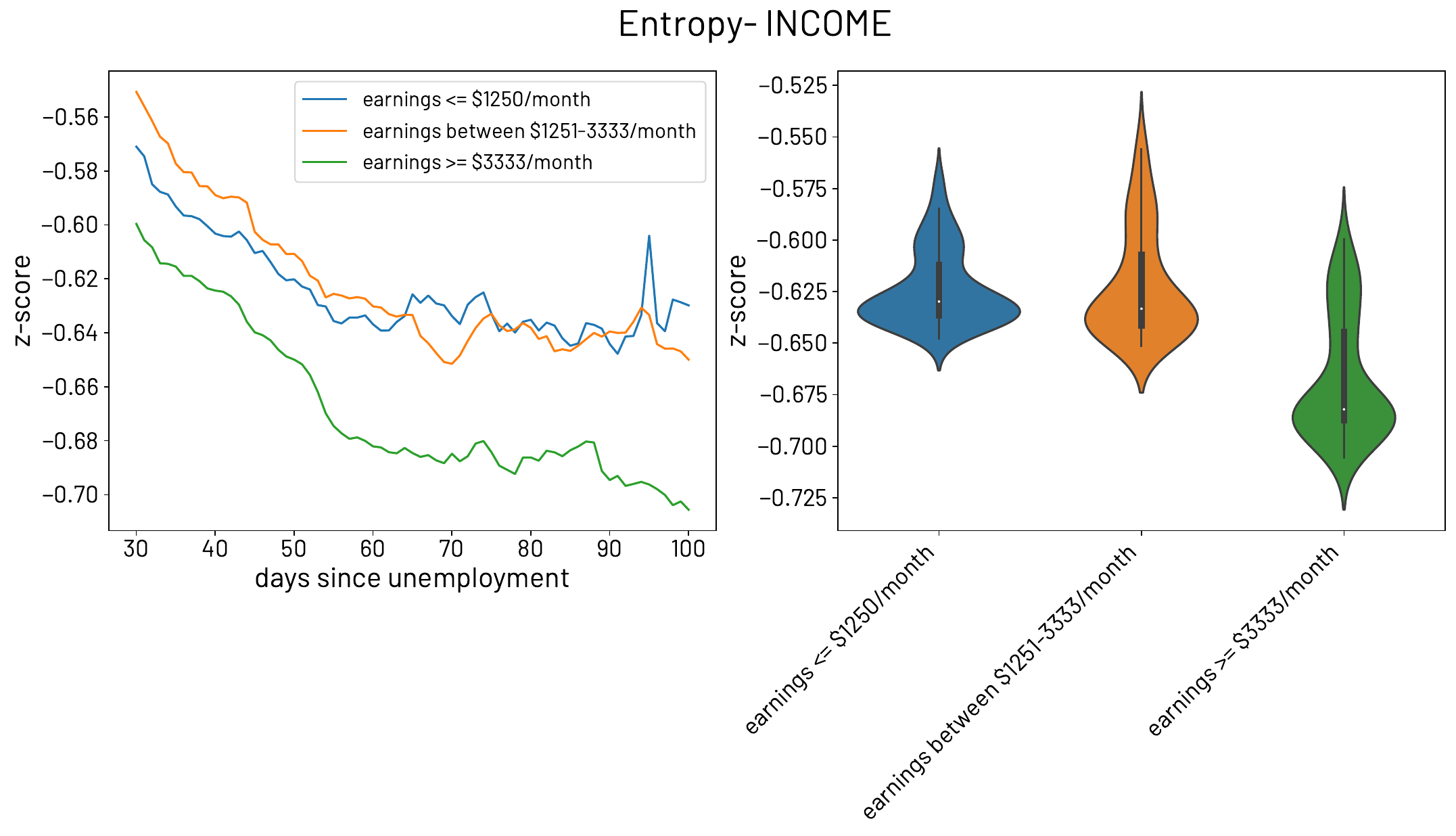}
\caption{Z-scores for the entropy divided by the Income demographic indicator and their corresponding distribution.}
\label{fig:decouple_income_entropy}
\end{figure}

\begin{figure}[!htb]
\centering
\includegraphics[width=\linewidth]{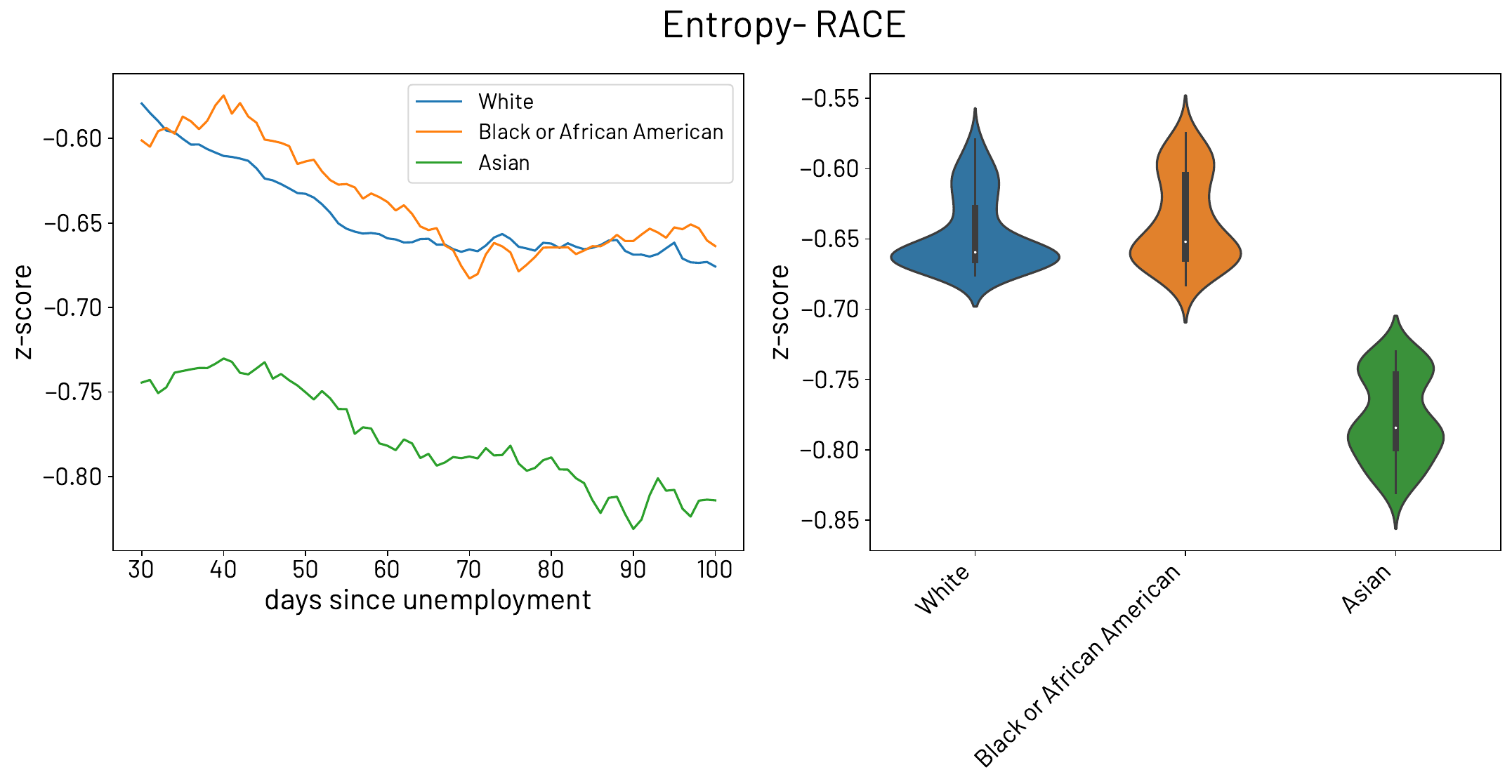}
\caption{Z-scores for the entropy divided by the Race demographic indicator and their corresponding distribution.}
\label{fig:decouple_race_entropy}
\end{figure}

\begin{figure}[!htb]
\centering
\includegraphics[width=\linewidth]{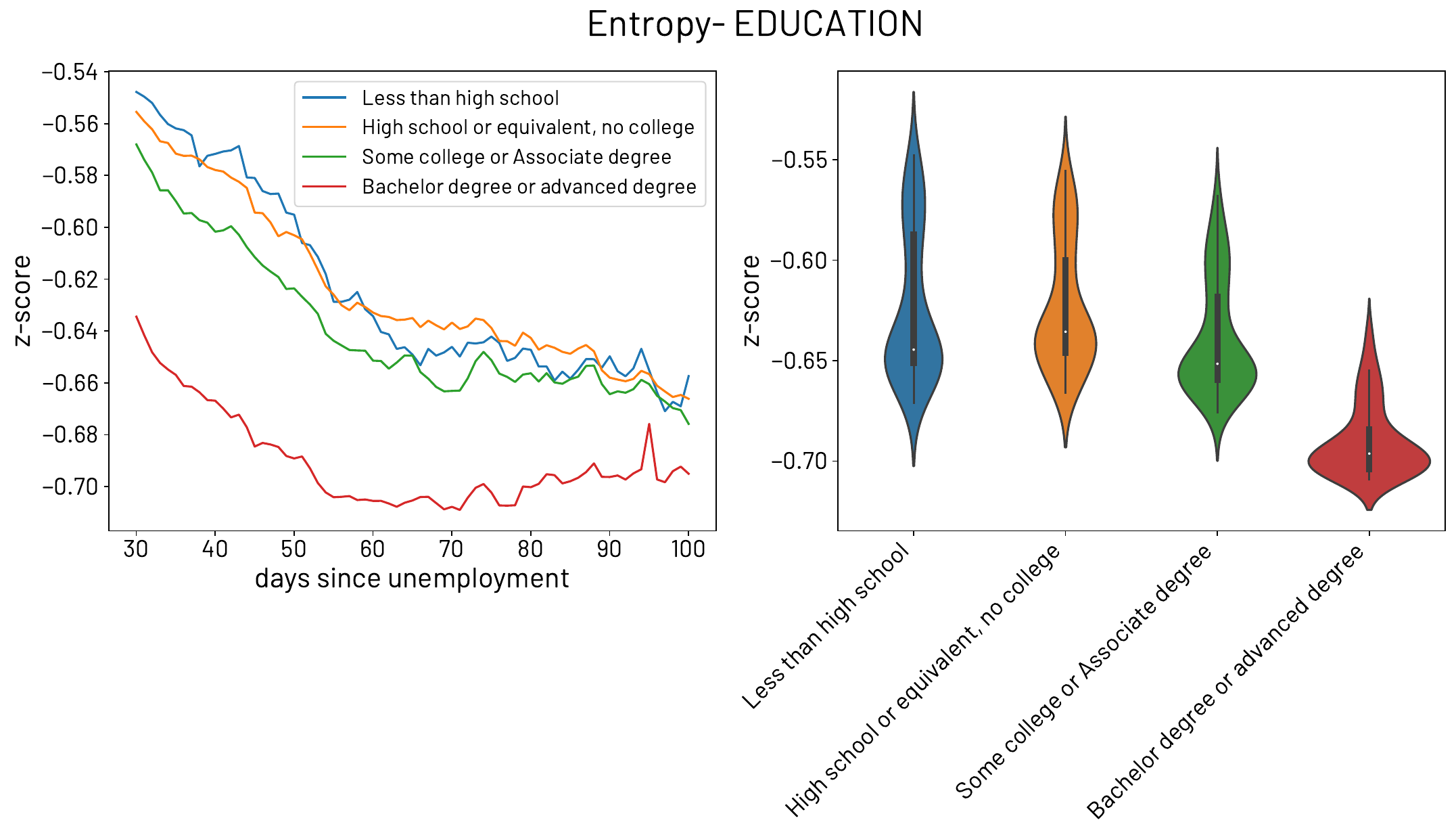}
\caption{Z-scores for the entropy divided by the Education demographic indicator and their corresponding distribution.}
\label{fig:decouple_education_entropy}
\end{figure}

\subsection{After job loss demographic differences: Capacity}\label{sec:capacity}

\begin{figure}[!htb]
\centering
\includegraphics[width=\linewidth]{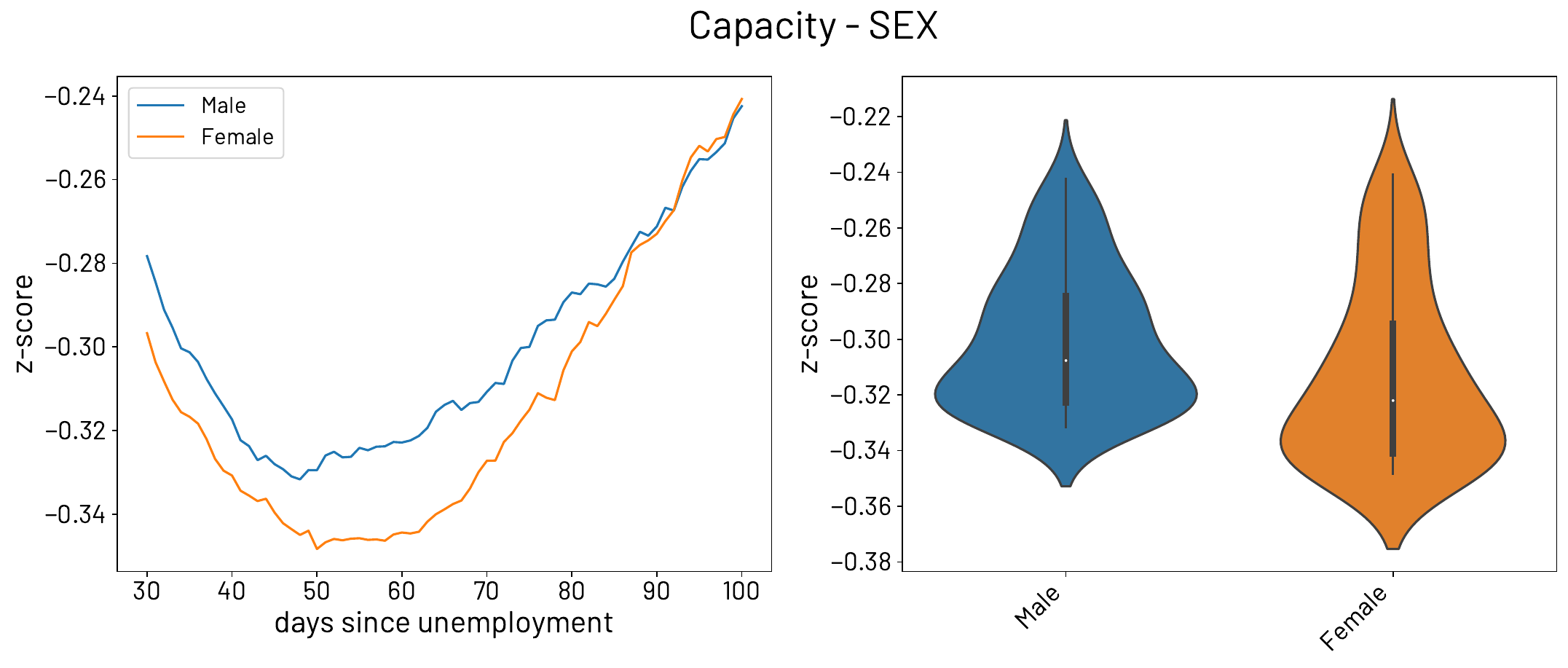}
\caption{Z-scores for the capacity divided by the Sex demographic indicator and their corresponding distribution.}
\label{fig:decouple_sex_capacity}
\end{figure}

\begin{figure}[!htb]
\centering
\includegraphics[width=\linewidth]{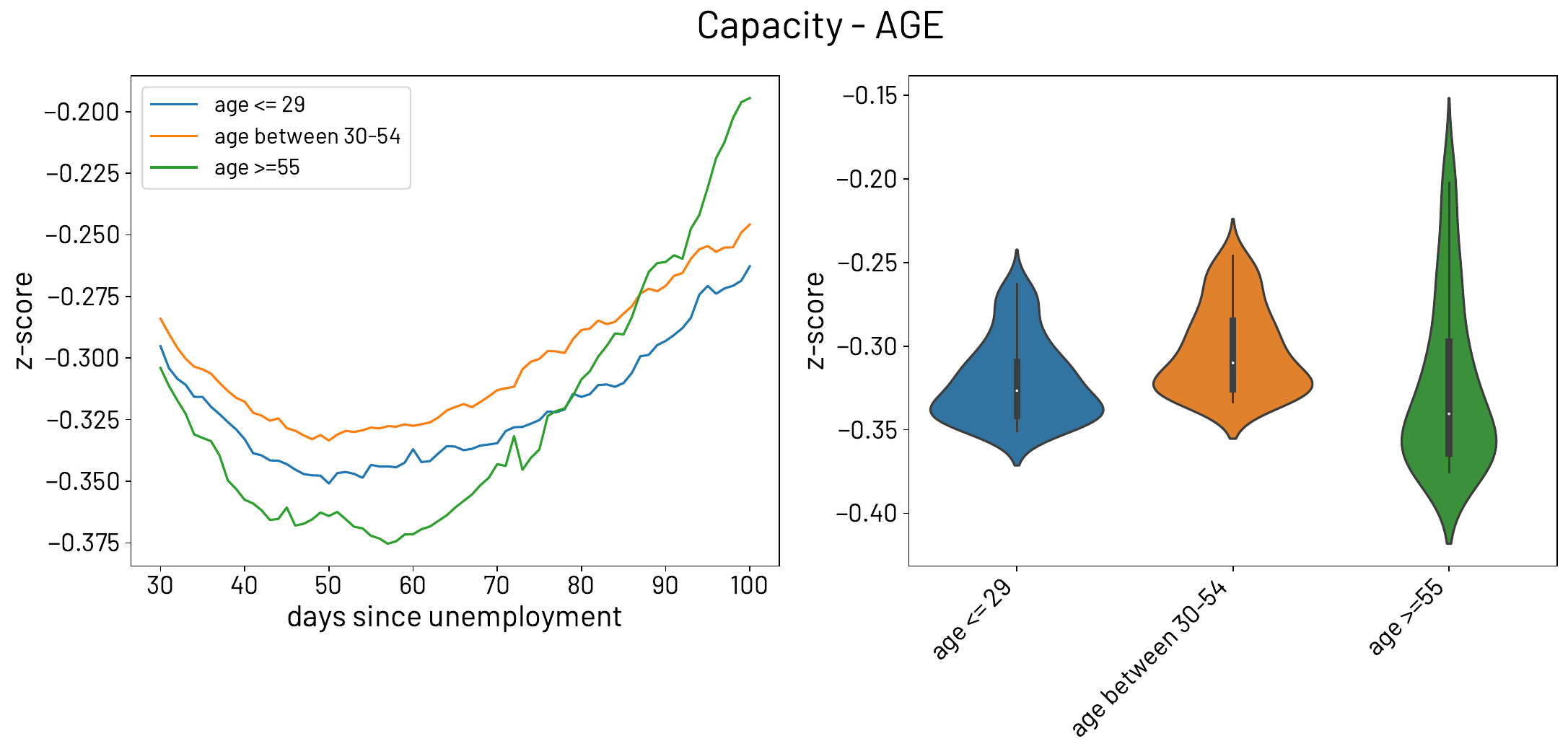}
\caption{Z-scores for the capacity divided by the Age demographic indicator and their corresponding distribution.}
\label{fig:decouple_age_capacity}
\end{figure}

\begin{figure}[!htb]
\centering
\includegraphics[width=\linewidth]{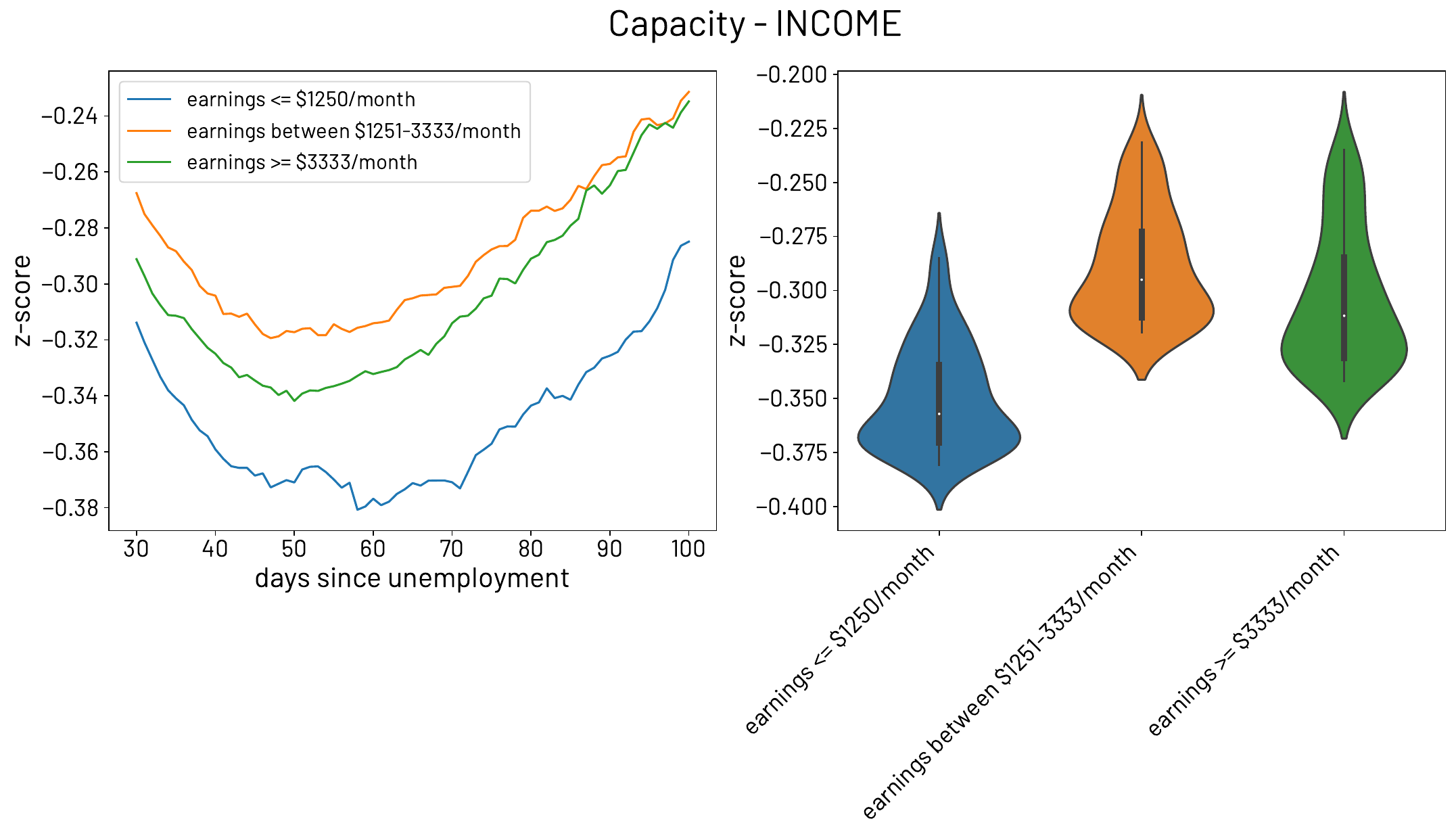}
\caption{Z-scores for the capacity divided by the Income demographic indicator and their corresponding distribution.}
\label{fig:decouple_income_capacity}
\end{figure}

\begin{figure}[!htb]
\centering
\includegraphics[width=\linewidth]{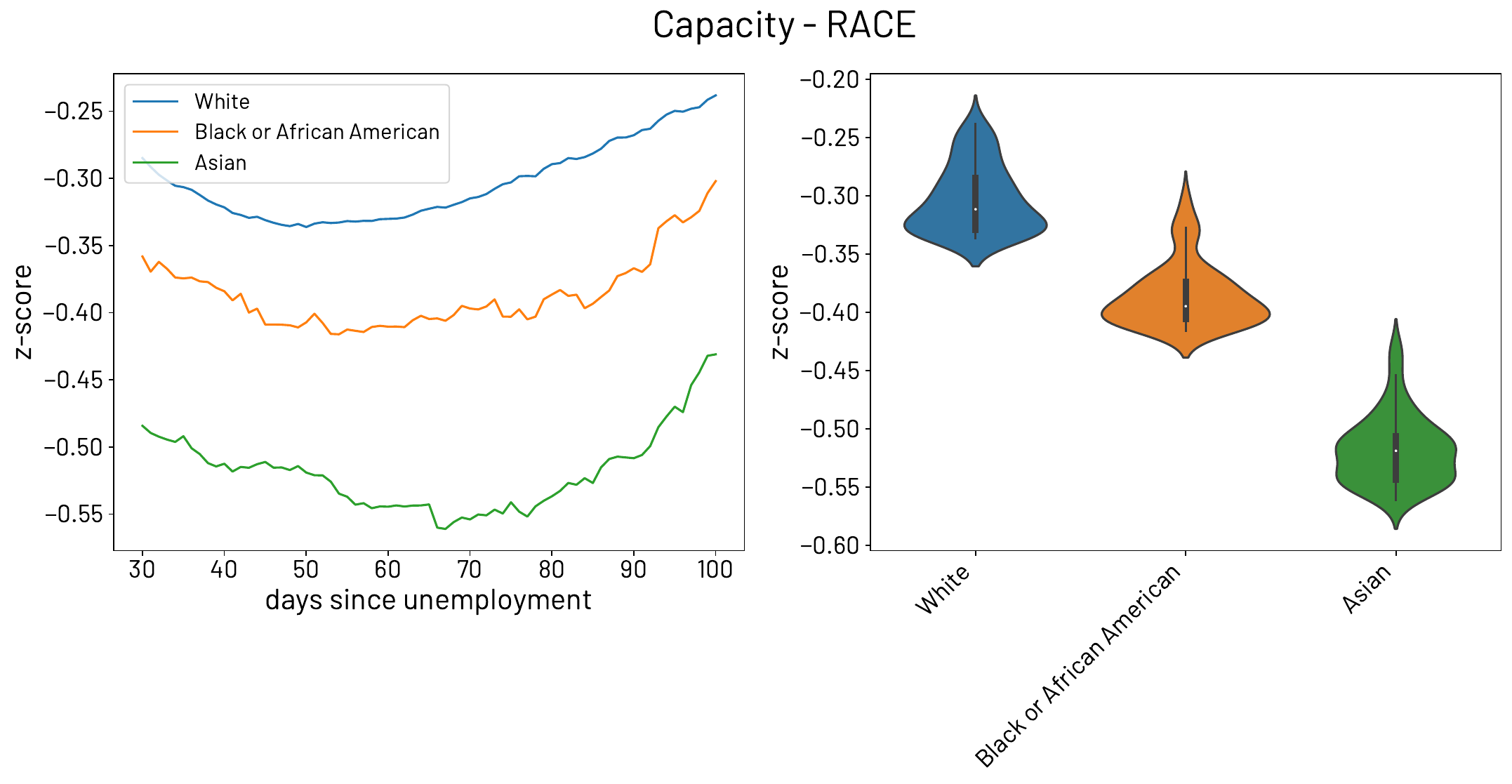}
\caption{Z-scores for the capacity divided by the Race demographic indicator and their corresponding distribution.}
\label{fig:decouple_race_capacity}
\end{figure}

\begin{figure}[!htb]
\centering
\includegraphics[width=\linewidth]{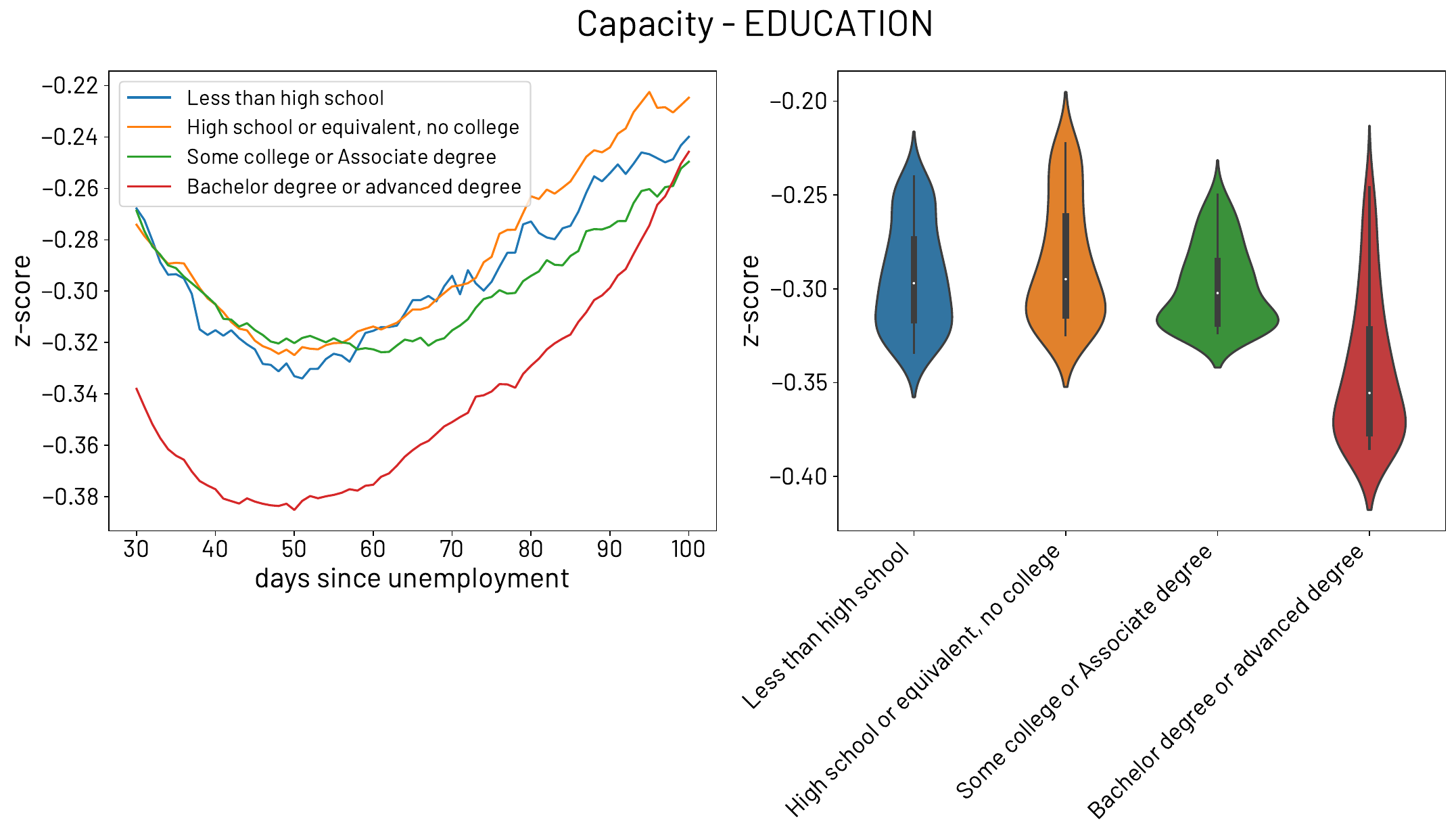}
\caption{Z-scores for the capacity divided by the Education demographic indicator and their corresponding distribution.}
\label{fig:decouple_education_capacity}
\end{figure}

\setlength{\tabcolsep}{9pt} 
\renewcommand{\arraystretch}{1} 
\begin{table}[]
\resizebox{\textwidth}{!}{%
\begin{tabular}{@{}lllrllr@{}}
\toprule
\textbf{Demographic} & \textbf{Mobility Metric} & \textbf{Group}                        & \textbf{Group avg} & \textbf{T-test} & \textbf{p-value} & \textbf{Significance} \\ \midrule
SEX       & Radius of gyration & Male                                  & -0.08 & -4.46e+01 & 0.00e+00  & *** \\
SEX       & Radius of gyration & Female                                & -0.1  & -6.47e+01 & 0.00e+00  & *** \\ \midrule
SEX       & Entropy            & Male                                  & -0.57 & -2.77e+02 & 0.00e+00  & *** \\
SEX       & Entropy            & Female                                & -0.6  & -3.14e+02 & 0.00e+00  & *** \\ \midrule
SEX       & Capacity           & Male                                  & -0.29 & -1.46e+02 & 0.00e+00  & *** \\
SEX       & Capacity           & Female                                & -0.31 & -1.57e+02 & 0.00e+00  & *** \\ \midrule
AGE       & Radius of gyration & age \textless{}= 29                   & -0.09 & -2.62e+01 & 1.87e-150 & *** \\
AGE       & Radius of gyration & age between 30-54                     & -0.09 & -6.37e+01 & 0.00e+00  & *** \\
AGE       & Radius of gyration & age \textgreater{}=55                 & -0.11 & -3.71e+01 & 3.78e-296 & *** \\ \midrule
AGE       & Entropy            & age \textless{}= 29                   & -0.59 & -1.52e+02 & 0.00e+00  & *** \\
AGE       & Entropy            & age between 30-54                     & -0.58 & -3.59e+02 & 0.00e+00  & *** \\
AGE       & Entropy            & age \textgreater{}=55                 & -0.59 & -1.48e+02 & 0.00e+00  & *** \\ \midrule
AGE       & Capacity           & age \textless{}= 29                   & -0.31 & -7.82e+01 & 0.00e+00  & *** \\
AGE       & Capacity           & age between 30-54                     & -0.29 & -1.84e+02 & 0.00e+00  & *** \\
AGE       & Capacity           & age \textgreater{}=55                 & -0.31 & -7.68e+01 & 0.00e+00  & *** \\ \midrule
INCOME    & Radius of gyration & earnings \textless{}= \$1250/month    & -0.09 & -2.96e+01 & 2.17e-191 & *** \\
INCOME    & Radius of gyration & earnings between \$1251-3333/month    & -0.1  & -4.33e+01 & 0.00e+00  & *** \\
INCOME    & Radius of gyration & earnings \textgreater{}= \$3333/month & -0.09 & -5.28e+01 & 0.00e+00  & *** \\ \midrule
INCOME    & Entropy            & earnings \textless{}= \$1250/month    & -0.57 & -1.54e+02 & 0.00e+00  & *** \\
INCOME    & Entropy            & earnings between \$1251-3333/month    & -0.56 & -2.03e+02 & 0.00e+00  & *** \\
INCOME    & Entropy            & earnings \textgreater{}= \$3333/month & -0.6  & -3.12e+02 & 0.00e+00  & *** \\ \midrule
INCOME    & Capacity           & earnings \textless{}= \$1250/month    & -0.33 & -8.84e+01 & 0.00e+00  & *** \\
INCOME    & Capacity           & earnings between \$1251-3333/month    & -0.28 & -1.10e+02 & 0.00e+00  & *** \\
INCOME    & Capacity           & earnings \textgreater{}= \$3333/month & -0.3  & -1.59e+02 & 0.00e+00  & *** \\ \midrule
RACE      & Radius of gyration & White                                 & -0.09 & -7.26e+01 & 0.00e+00  & *** \\
RACE      & Radius of gyration & Black or African American             & -0.12 & -1.31e+01 & 1.49e-38  & *** \\
RACE      & Radius of gyration & Asian                                 & -0.13 & -1.08e+01 & 1.12e-26  & *** \\ \midrule
RACE      & Entropy            & White                                 & -0.58 & -3.95e+02 & 0.00e+00  & *** \\
RACE      & Entropy            & Black or African American             & -0.61 & -5.06e+01 & 0.00e+00  & *** \\
RACE      & Entropy            & Asian                                 & -0.74 & -5.61e+01 & 0.00e+00  & *** \\ \midrule
RACE      & Capacity           & White                                 & -0.29 & -2.02e+02 & 0.00e+00  & *** \\
RACE      & Capacity           & Black or African American             & -0.4  & -3.77e+01 & 6.53e-276 & *** \\
RACE      & Capacity           & Asian                                 & -0.52 & -4.38e+01 & 1.76e-320 & *** \\ \midrule
EDUCATION & Radius of gyration & Less than high school                 & -0.1  & -1.82e+01 & 1.51e-73  & *** \\
EDUCATION            & Radius of gyration       & High school or equivalent, no college & -0.09              & -3.83e+01       & 0.00e+00         & ***                   \\
EDUCATION & Radius of gyration & Some college or Associate degree      & -0.09 & -4.67e+01 & 0.00e+00  & *** \\
EDUCATION & Radius of gyration & Bachelor degree or advanced degree    & -0.09 & -3.97e+01 & 0.00e+00  & *** \\ \midrule
EDUCATION & Entropy            & Less than high school                 & -0.56 & -8.35e+01 & 0.00e+00  & *** \\
EDUCATION & Entropy            & High school or equivalent, no college & -0.56 & -1.95e+02 & 0.00e+00  & *** \\
EDUCATION & Entropy            & Some college or Associate degree      & -0.57 & -2.53e+02 & 0.00e+00  & *** \\
EDUCATION & Entropy            & Bachelor degree or advanced degree    & -0.63 & -2.37e+02 & 0.00e+00  & *** \\ \midrule
EDUCATION & Capacity           & Less than high school                 & -0.29 & -4.53e+01 & 0.00e+00  & *** \\
EDUCATION & Capacity           & High school or equivalent, no college & -0.28 & -9.92e+01 & 0.00e+00  & *** \\
EDUCATION & Capacity           & Some college or Associate degree      & -0.28 & -1.24e+02 & 0.00e+00  & *** \\
EDUCATION & Capacity           & Bachelor degree or advanced degree    & -0.33 & -1.20e+02 & 0.00e+00  & *** \\ \bottomrule
\end{tabular}%
}
\caption{Comparison of the groups of a socio-demographic factor against the entire employed population. The statistical significance is computed with the T-test statistic.}
\label{tab:unemployed_vs_all_empl_population}
\end{table}

\begin{table}[]
\resizebox{\textwidth}{!}{%
\begin{tabular}{@{}llllrlr@{}}
\toprule
\textbf{Demographic} &
  \textbf{Mobility Metric} &
  \textbf{Group1} &
  \textbf{Group2} &
  \textbf{Welch t-test} &
  \textbf{p-value} &
  \textbf{Significance} \\ \midrule
SEX       & Radius of gyration & Male                                  & Female                                & 10.13  & 3.90e-24 & *** \\
SEX       & Entropy            & Male                                  & Female                                & 9.96   & 2.22e-23 & *** \\
SEX       & Capacity           & Male                                  & Female                                & 6.21   & 5.39e-10 & *** \\ \midrule
AGE       & Radius of gyration & age \textless{}= 29                   & age between 30-54                     & -1.06  & 2.89e-01 & ns  \\
AGE       & Radius of gyration & age \textless{}= 29                   & age \textgreater{}=55                 & 4.08   & 4.59e-05 & *** \\
AGE       & Radius of gyration & age between 30-54                     & age \textgreater{}=55                 & 6.92   & 4.53e-12 & *** \\ \midrule
AGE       & Entropy            & age \textless{}= 29                   & age between 30-54                     & -3.10  & 1.94e-03 & **  \\
AGE       & Entropy            & age \textless{}= 29                   & age \textgreater{}=55                 & -0.53  & 5.94e-01 & ns  \\
AGE       & Entropy            & age between 30-54                     & age \textgreater{}=55                 & 2.36   & 1.85e-02 & *   \\ \midrule
AGE       & Capacity           & age \textless{}= 29                   & age between 30-54                     & -3.31  & 9.48e-04 & *** \\
AGE       & Capacity           & age \textless{}= 29                   & age \textgreater{}=55                 & -0.25  & 8.02e-01 & ns  \\
AGE       & Capacity           & age between 30-54                     & age \textgreater{}=55                 & 2.94   & 3.29e-03 & **  \\ \midrule
INCOME    & Radius of gyration & earnings \textless{}= \$1250/month    & earnings between \$1251-3333/month    & 2.04   & 4.10e-02 & *   \\
INCOME    & Radius of gyration & earnings \textless{}= \$1250/month    & earnings \textgreater{}= \$3333/month & -0.80  & 4.21e-01 & ns  \\
INCOME    & Radius of gyration & earnings between \$1251-3333/month    & earnings \textgreater{}= \$3333/month & -3.76  & 1.71e-04 & *** \\ \midrule
INCOME    & Entropy            & earnings \textless{}= \$1250/month    & earnings between \$1251-3333/month    & -1.48  & 1.39e-01 & ns  \\
INCOME    & Entropy            & earnings \textless{}= \$1250/month    & earnings \textgreater{}= \$3333/month & 9.60   & 8.40e-22 & *** \\
INCOME    & Entropy            & earnings between \$1251-3333/month    & earnings \textgreater{}= \$3333/month & 13.84  & 1.61e-43 & *** \\ \midrule
INCOME    & Capacity           & earnings \textless{}= \$1250/month    & earnings between \$1251-3333/month    & -9.93  & 3.29e-23 & *** \\
INCOME    & Capacity           & earnings \textless{}= \$1250/month    & earnings \textgreater{}= \$3333/month & -7.49  & 7.04e-14 & *** \\
INCOME    & Capacity           & earnings between \$1251-3333/month    & earnings \textgreater{}= \$3333/month & 4.32   & 1.55e-05 & *** \\ \midrule
RACE      & Radius of gyration & White                                 & Black or African American             & 2.98   & 2.92e-03 & **  \\
RACE      & Radius of gyration & White                                 & Asian                                 & 3.54   & 4.03e-04 & *** \\
RACE      & Radius of gyration & Black or African American             & Asian                                 & 1.14   & 2.55e-01 & ns  \\ \midrule
RACE      & Entropy            & White                                 & Black or African American             & 1.95   & 5.16e-02 & ns  \\
RACE      & Entropy            & White                                 & Asian                                 & 11.99  & 2.63e-32 & *** \\
RACE      & Entropy            & Black or African American             & Asian                                 & 7.63   & 2.77e-14 & *** \\ \midrule
RACE      & Capacity           & White                                 & Black or African American             & 10.25  & 1.89e-24 & *** \\
RACE      & Capacity           & White                                 & Asian                                 & 19.18  & 2.44e-77 & *** \\
RACE      & Capacity           & Black or African American             & Asian                                 & 7.50   & 7.34e-14 & *** \\ \midrule
EDUCATION & Radius of gyration & Less than high school                 & High school or equivalent, no college & -1.14  & 2.55e-01 & ns  \\
EDUCATION & Radius of gyration & Less than high school                 & Some college or Associate degree      & -1.50  & 1.34e-01 & ns  \\
EDUCATION & Radius of gyration & Less than high school                 & Bachelor degree or advanced degree    & -0.68  & 4.98e-01 & ns  \\
EDUCATION & Radius of gyration & High school or equivalent, no college & Some college or Associate degree      & -0.61  & 5.41e-01 & ns  \\
EDUCATION &
  Radius of gyration &
  High school or equivalent, no college &
  Bachelor degree or advanced degree &
  0.81 &
  4.20e-01 &
  ns \\
EDUCATION & Radius of gyration & Some college or Associate degree      & Bachelor degree or advanced degree    & 1.50   & 1.33e-01 & ns  \\ \midrule
EDUCATION & Entropy            & Less than high school                 & High school or equivalent, no college & 0.54   & 5.86e-01 & ns  \\
EDUCATION & Entropy            & Less than high school                 & Some college or Associate degree      & 2.22   & 2.65e-02 & *   \\
EDUCATION & Entropy            & Less than high school                 & Bachelor degree or advanced degree    & 10.43  & 1.98e-25 & *** \\
EDUCATION & Entropy            & High school or equivalent, no college & Some college or Associate degree      & 3.18   & 1.46e-03 & **  \\
EDUCATION & Entropy            & High school or equivalent, no college & Bachelor degree or advanced degree    & 18.08  & 5.42e-73 & *** \\
EDUCATION & Entropy            & Some college or Associate degree      & Bachelor degree or advanced degree    & 16.96  & 1.76e-64 & *** \\ \midrule
EDUCATION & Capacity           & Less than high school                 & High school or equivalent, no college & -1.82  & 6.85e-02 & ns  \\
EDUCATION & Capacity           & Less than high school                 & Some college or Associate degree      & -1.15  & 2.51e-01 & ns  \\
EDUCATION & Capacity           & Less than high school                 & Bachelor degree or advanced degree    & 5.59   & 2.25e-08 & *** \\
EDUCATION & Capacity           & High school or equivalent, no college & Some college or Associate degree      & 1.37   & 1.72e-01 & ns  \\
EDUCATION & Capacity           & High school or equivalent, no college & Bachelor degree or advanced degree    & 13.22  & 7.47e-40 & *** \\
EDUCATION & Capacity           & Some college or Associate degree      & Bachelor degree or advanced degree    & 13.14  & 2.07e-39 & *** \\ \bottomrule
\end{tabular}%
}
\caption{Comparison to test whether there are differences between groups inside the same socio-demographic factor (e.g., male unemployed vs female unemployed). The statistical significance is computed with the Welch T-test statistic.}
\label{tab:unemployed_within_demographic_group}
\end{table}

\begin{table}[]
\resizebox{0.95\textwidth}{!}{%
\begin{tabular}{@{}lllrrrr@{}}
\toprule
\textbf{Demographic} & \textbf{Mobility Metric} & \textbf{Group}                        & \textbf{Group avg} & \textbf{T-test} & \textbf{p-value} & \textbf{Significance} \\ \midrule
SEX       & Radius of gyration & Male                                  & -0.09 & -4.98e+01 & 0.00e+00  & *** \\
SEX       & Radius of gyration & Female                                & -0.09 & -5.81e+01 & 0.00e+00  & *** \\ \midrule
SEX       & Entropy            & Male                                  & -0.59 & -2.90e+02 & 0.00e+00  & *** \\
SEX       & Entropy            & Female                                & -0.59 & -3.11e+02 & 0.00e+00  & *** \\ \midrule
SEX       & Capacity           & Male                                  & -0.3  & -1.53e+02 & 0.00e+00  & *** \\
SEX       & Capacity           & Female                                & -0.29 & -1.43e+02 & 0.00e+00  & *** \\ \midrule
AGE       & Radius of gyration & age \textless{}= 29                   & -0.09 & -2.45e+01 & 1.48e-131 & *** \\
AGE       & Radius of gyration & age between 30-54                     & -0.09 & -6.37e+01 & 0.00e+00  & *** \\
AGE       & Radius of gyration & age \textgreater{}=55                 & -0.08 & -2.05e+01 & 4.24e-93  & *** \\ \midrule
AGE       & Entropy            & age \textless{}= 29                   & -0.63 & -1.61e+02 & 0.00e+00  & *** \\
AGE       & Entropy            & age between 30-54                     & -0.58 & -3.61e+02 & 0.00e+00  & *** \\
AGE       & Entropy            & age \textgreater{}=55                 & -0.62 & -1.50e+02 & 0.00e+00  & *** \\ \midrule
AGE       & Capacity           & age \textless{}= 29                   & -0.31 & -7.75e+01 & 0.00e+00  & *** \\
AGE       & Capacity           & age between 30-54                     & -0.29 & -1.77e+02 & 0.00e+00  & *** \\
AGE       & Capacity           & age \textgreater{}=55                 & -0.32 & -7.96e+01 & 0.00e+00  & *** \\ \midrule
INCOME    & Radius of gyration & earnings \textless{}= \$1250/month    & -0.09 & -2.79e+01 & 1.60e-169 & *** \\
INCOME    & Radius of gyration & earnings between \$1251-3333/month    & -0.08 & -3.38e+01 & 4.20e-249 & *** \\
INCOME    & Radius of gyration & earnings \textgreater{}= \$3333/month & -0.09 & -5.78e+01 & 0.00e+00  & *** \\ \midrule
INCOME    & Entropy            & earnings \textless{}= \$1250/month    & -0.59 & -1.70e+02 & 0.00e+00  & *** \\
INCOME    & Entropy            & earnings between \$1251-3333/month    & -0.58 & -2.27e+02 & 0.00e+00  & *** \\
INCOME    & Entropy            & earnings \textgreater{}= \$3333/month & -0.59 & -3.16e+02 & 0.00e+00  & *** \\ \midrule
INCOME    & Capacity           & earnings \textless{}= \$1250/month    & -0.32 & -8.11e+01 & 0.00e+00  & *** \\
INCOME    & Capacity           & earnings between \$1251-3333/month    & -0.27 & -1.04e+02 & 0.00e+00  & *** \\
INCOME    & Capacity           & earnings \textgreater{}= \$3333/month & -0.3  & -1.63e+02 & 0.00e+00  & *** \\ \midrule
RACE      & Radius of gyration & White                                 & -0.09 & -7.56e+01 & 0.00e+00  & *** \\
RACE      & Radius of gyration & Black or African American             & 0.13  & 3.03e+00  & 2.48e-03  & **  \\
RACE      & Radius of gyration & Asian                                 & 0.76  & 3.21e+00  & 1.37e-03  & **  \\ \midrule
RACE      & Entropy            & White                                 & -0.59 & -4.19e+02 & 0.00e+00  & *** \\
RACE      & Entropy            & Black or African American             & -0.58 & -5.07e+01 & 0.00e+00  & *** \\
RACE      & Entropy            & Asian                                 & -0.62 & -4.41e+01 & 2.96e-323 & *** \\ \midrule
RACE      & Capacity           & White                                 & -0.3  & -2.07e+02 & 0.00e+00  & *** \\
RACE      & Capacity           & Black or African American             & -0.29 & -2.56e+01 & 9.35e-137 & *** \\
RACE      & Capacity           & Asian                                 & -0.32 & -1.75e+01 & 1.26e-65  & *** \\ \midrule
EDUCATION & Radius of gyration & Less than high school                 & -0.05 & -5.97e+00 & 2.39e-09  & *** \\
EDUCATION            & Radius of gyration       & High school or equivalent, no college & -0.09              & -3.67e+01       & 4.85e-292        & ***                   \\
EDUCATION & Radius of gyration & Some college or Associate degree      & -0.09 & -4.81e+01 & 0.00e+00  & *** \\
EDUCATION & Radius of gyration & Bachelor degree or advanced degree    & -0.09 & -4.34e+01 & 0.00e+00  & *** \\ \midrule
EDUCATION & Entropy            & Less than high school                 & -0.6  & -9.78e+01 & 0.00e+00  & *** \\
EDUCATION & Entropy            & High school or equivalent, no college & -0.59 & -2.15e+02 & 0.00e+00  & *** \\
EDUCATION & Entropy            & Some college or Associate degree      & -0.57 & -2.66e+02 & 0.00e+00  & *** \\
EDUCATION & Entropy            & Bachelor degree or advanced degree    & -0.61 & -2.37e+02 & 0.00e+00  & *** \\ \midrule
EDUCATION & Capacity           & Less than high school                 & -0.3  & -4.73e+01 & 0.00e+00  & *** \\
EDUCATION & Capacity           & High school or equivalent, no college & -0.3  & -1.12e+02 & 0.00e+00  & *** \\
EDUCATION & Capacity           & Some college or Associate degree      & -0.29 & -1.30e+02 & 0.00e+00  & *** \\
EDUCATION & Capacity           & Bachelor degree or advanced degree    & -0.3  & -1.09e+02 & 0.00e+00  & *** \\ \bottomrule
\end{tabular}%
}
\caption{Comparison of the groups of a socio-demographic factor against their corresponding employed population (e.g., male unemployed vs male employed). The statistical significance is computed with the T-test statistic.}
\label{tab:unempl_empl_same_demographic_group}
\end{table}